\documentclass{aa}

\usepackage[varg]{txfonts}
\usepackage{graphicx}
\usepackage[usenames,dvipsnames]{color}

\usepackage{natbib} \bibpunct{(}{)}{;}{a}{}{,}

\newcommand{\bvp}{}
\newcommand*{\factor}{0.99} 
\newcommand*{\vfac}{-0.1cm}
\newcommand*{\spfactor}{1.00cm}
\newcommand{\vesc}{$\varv_{\rm esc}$\,}
\newcommand{\vinf}{$\varv_{\infty}$}
\begin{document}

\title{On the H$\alpha$ Behaviour of Blue Supergiants: Rise and Fall over the Bi-stability Jump}

\author{Blagovest Petrov  \and Jorick S. Vink \and G\"{o}tz Gr\"{a}fener}

\institute{Armagh Observatory, College Hill, Armagh, BT61 9DG, Northern Ireland}

\date{Received date:  / Accepted date: }

\abstract    
  {The evolutionary state of blue supergiants is still unknown. Stellar wind mass loss is one of the dominant processes determining the evolution of massive stars, and it may provide clues on the evolutionary properties of blue supergiants. As the H$\alpha$ line is the most oft-used mass-loss tracer in the OB-star regime, we investigate H$\alpha$ line formation as a function of $T_{\rm eff}$.}
  {We provide a detailed analysis of the H$\alpha$ line for OB supergiant models over an $T_{\rm eff}$ range between 30\,000 and 12\,500\,K, with the aim of understanding the mass-loss properties of blue supergiants.}
  {We model the H$\alpha$ line using the non-LTE code {\sc cmfgen}, in the context of the bi-stability jump at $T_{\rm eff}$ $\sim$ 22\,500\,K.}
  {We find a maximum in the H$\alpha$ equivalent width at 22\,500\,K -  \textit{exactly} at the location of the bi-stability jump. The H$\alpha$ line-profile behaviour is characterised by two branches of effective temperature: (i) a  \textit{hot branch} between 30\,000 and 22\,500\,K, where H$\alpha$ emission becomes stronger with decreasing $T_{\rm eff}$, and (ii) a \textit{cool branch} between 22\,500 and 12\,500\,K, where the H$\alpha$ line becomes \textit{weaker}. Our models show that this non-monotonic H$\alpha$ behaviour is related to the optical depth of Ly$\alpha$, finding that at the ``cool'' branch the population of the 2$^{\rm nd}$ level of hydrogen is enhanced in comparison to the 3$^{\rm rd}$level. This is expected to increase line absorption, leading to weaker H$\alpha$ flux when $T_{\rm eff}$ drops from 22\,500\,K downwards. We also show that for late B supergiants (at $T_{\rm eff}$ below $\sim$15\,000\,K), the differences in the H$\alpha$ line between homogeneous and clumpy winds becomes insignificant. Moreover, we show that at the bi-stability jump H$\alpha$ changes its character completely, from an optically thin to an optically thick line, implying that \textit{macro-clumping} should play an important role at temperatures below the bi-stability jump. This would not only have consequences for the character of observed H$\alpha$ line profiles, but also for the reported discrepancies between theoretical and empirical mass-loss rates.}{} 
\titlerunning{On the H$\alpha$ behaviour for BSGs}
\authorrunning{B. Petrov \it{et al.}}

\maketitle

\vspace{\vfac}
\section{Introduction}

  Blue supergiants (BSG) are key creators of heavy elements, thereby contributing to the composition of the interstellar medium; they are visible out to large distances, and dominate the spectra of star-formation galaxies. Moreover, they heat the interstellar gas and dust and produce the far-infrared luminosities of galaxies. B supergiants (Bsgs) in particular are indissoluble troublemakers in today's Astrophysics, as a proper understanding of their evolution is still in its infancy \citep{langer12}.  Even their fundamental properties are not yet known, and the basic issue as to whether they are core hydrogen (H) burning main-sequence (MS) or core helium (He) burning post-MS stars is still under debate \citep{vink10,georgy13}.

  \cite{lam95} reported the existence of a drop in terminal wind velocities of Bsgs by a factor of two, which is referred to as the ``bi-stability jump'' \citep{pauldrach90}. According to model predictions of \citet{vink99} it is expected that the faster winds of hotter stars switch to slower winds with a $\sim$ 5 times higher $\dot{M}$ below $T_{\rm eff}$ $\sim 22\,000$\,K. Whereas the drop in terminal velocities has been confirmed by observations \citep{markova08}, the predicted increase in $\dot{M}$ is still controversial. 

  Whilst \cite{ben07} and \cite{markova08} uncovered a local maximum in both radio and H$\alpha$ mass-loss rates at the location of the bi-stability jump, several works have highlighted significant discrepancies between theoretical and empirical mass-loss rates for B1 and later B-type supergiants \citep{vink00,trundle04,trundle05,cro06,ben07,markova08}. For the late Bsgs \cite{vink00} noted that their predictions agreed reasonably well with empirical rates from both radio and H$\alpha$ emission, but they found huge (order of magnitude) discrepancies when the H$\alpha$ line was P\,Cyg shaped or in absorption. \cite{searle08} emphasised that empirical models for Bsgs likely have an incorrect ionisation structure as they found it challenging to reproduce the optical H$\alpha$ line simultaneously with key ultraviolet (UV) diagnostics.

  To make the picture even more complex, one should be aware of discrepancies between mass-loss rates estimated from H$\alpha$, UV, and radio observations for OB stars in general \citep{massa03,bouret05,puls06,full06}, which may be due to distance-dependent wind clumping and/or porosity effects \citep{Oskinova07,Sundq10,sundqvist11,Muijres11,Surlan12}.  

  In any case, the general trend seems to indicate $\dot{M}_{vink} > \dot{M}_{\rm{H}_{\alpha}}$ for B1 and later supergiants, whilst the reverse holds for  earlier O supergiants.  The key question is whether this discrepancy is the result of incorrect predictions or alternatively that we may not understand the mass-loss indicator H$\alpha$ well enough. What would one expect to happen when Fe\begin{small}IV\end{small} recombines, and Fe\begin{small}III\end{small} starts to control the wind driving \citep{vink99}?

  Over the last decade we have made a dedicated effort to improve the physics in the Monte Carlo line driving calculations.  We now solve the wind dynamics more locally consistently \citep{muller08,muijres12}, we added Fe to the statistical equilibrium calculations in the {\sc isa-wind} model atmosphere \citep{dekoter93}(rather than treating this important line-driving element in a modified nebular approximation), and finally, we have studied the effects of wind clumping and porosity on the line driving \citep{Muijres11}. After all these improvements in the Monte Carlo line-driving physics, we have to admit that the basic problem of $\dot{M}_{vink} > \dot{M}_{\rm{H}_{\alpha}}$ for B1 and later supergiants, is still present, and it is time that we also consider the possibility that it is not the predictions that are at fault, but that we do not understand H$\alpha$ line formation in Bsgs sufficiently well to allow accurate mass-loss determinations from  H$\alpha$.

  In order to obtain a more complete picture of H$\alpha$ as a mass-loss diagnostic, we need to know not only the clumping properties of BSGs, but also how sensitive H$\alpha$ is to the clumping  on  both sides of the bi-stability jump. Currently, $T_{\rm eff}$ dependence of the  mass-loss rates from H$\alpha$ of BSGs is still uncertain, and a better knowledge is required, especially as it impacts the question of the evolutionary nature of BSGs.

  \begin{figure*}
      \centering
      \resizebox{\factor\hsize}{!}{\includegraphics{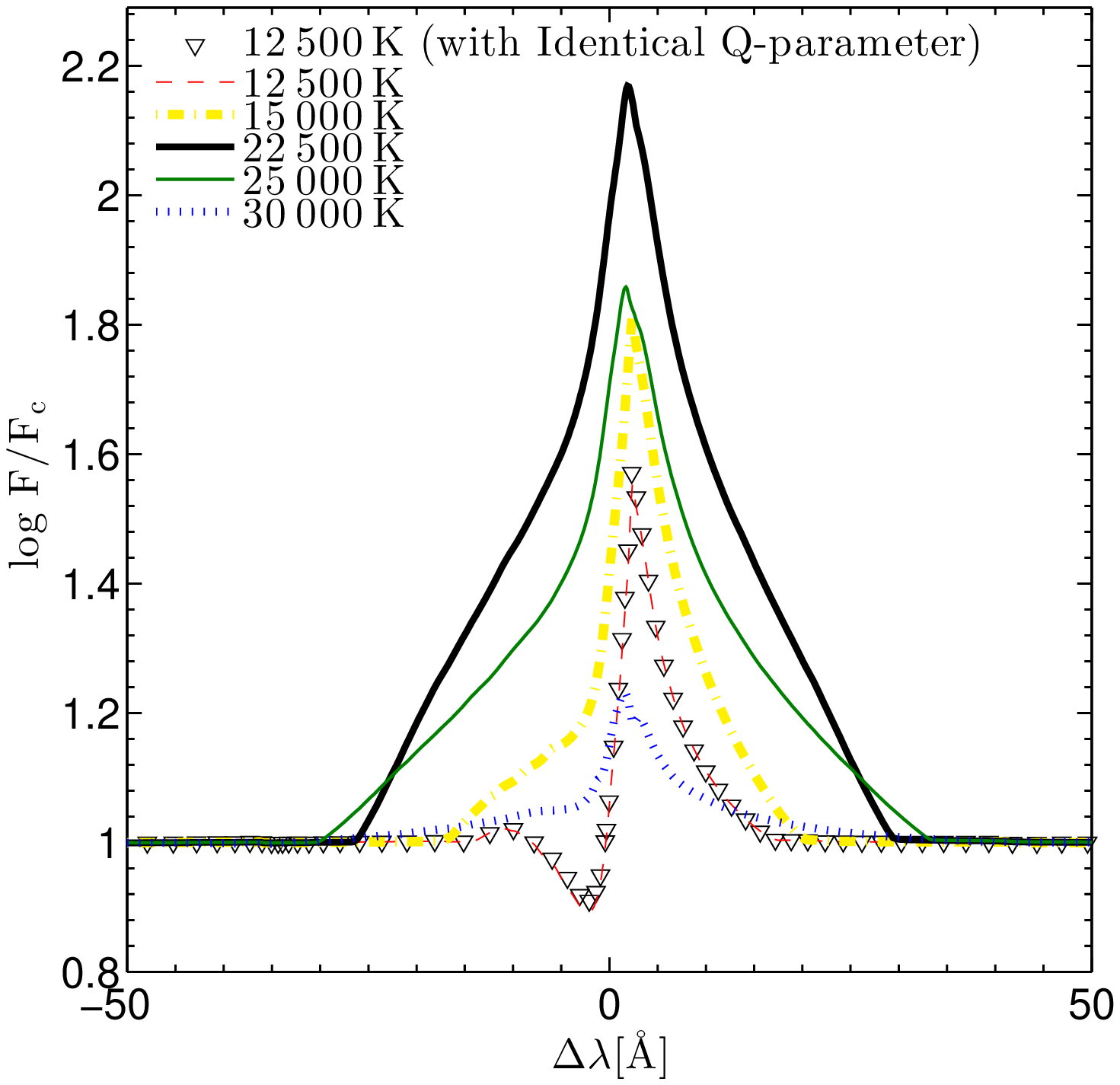}\hspace{\spfactor}  \includegraphics{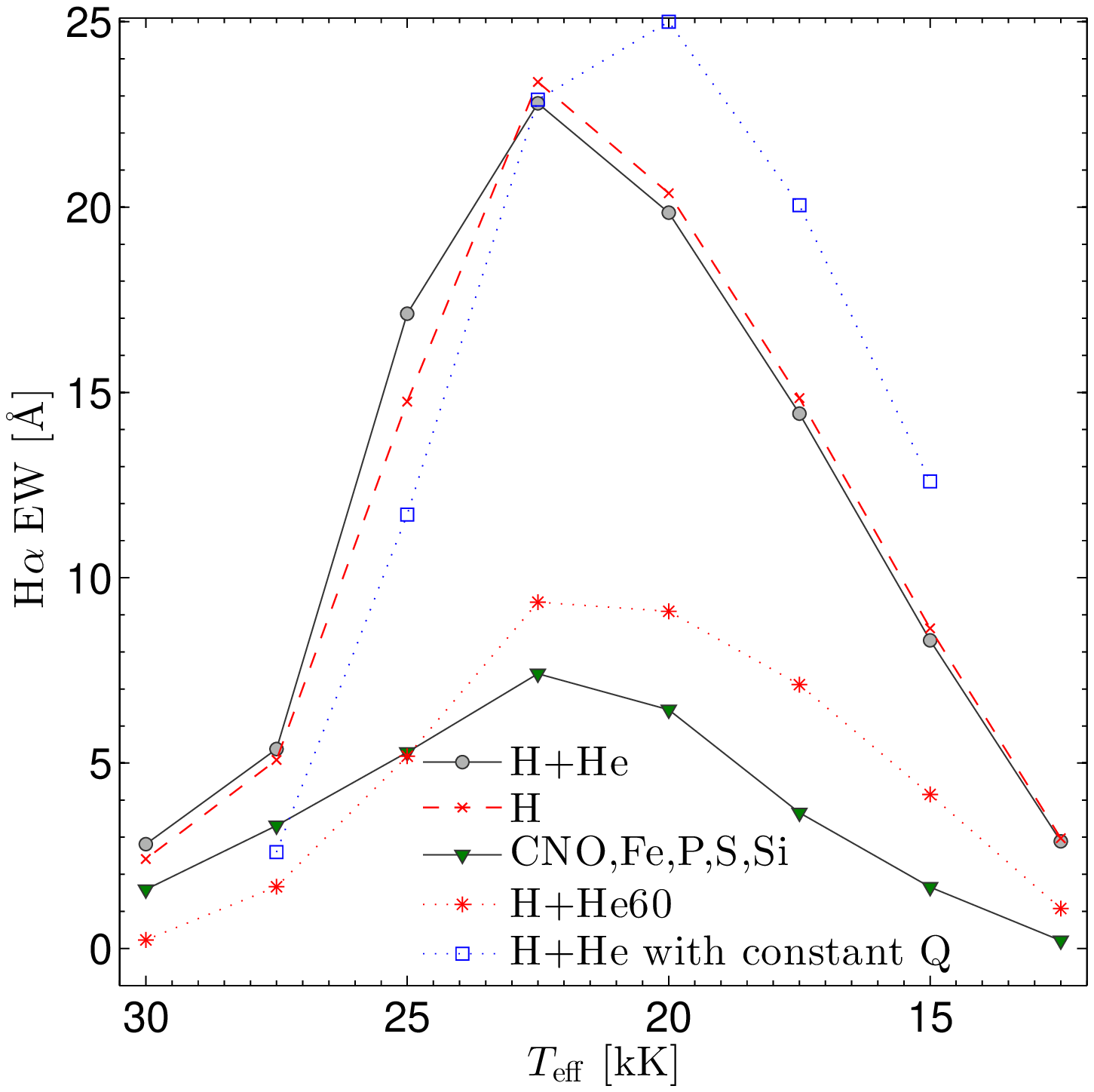}}
      \caption{\textit{Left:} H${\alpha}$ line profiles for {\sc cmfgen} models   with parameters as listed in Table~\ref{table:tab1}. {\bvp Black triangles}   represent the line profile with the same $Q$-parameter (Eq.\,\ref{eq_Q})   as model C ($T_{\rm eff}$=12\,500\,K) but with different $\dot{M}$ and $R_{\star}$ values. \textit{Right:} H$\alpha$ line EW vs $T_{\rm eff}$ for models with only H (crosses), H+He (circles) and more sophisticated (triangles) models. {\bvp Red asterisks} represent the changes in the H$\alpha$ line when the He mass fraction in the pure H+He models  is  increased to 60\%. Blue squares indicate how the H$\alpha$ EW behaves as a   function of a constant $Q$ value.}
      \label{fig:HaLP}
  \end{figure*}

  Despite the fact that the spectral modelling of Bsgs is an active area of research \citep{zorec09,fraser10,castro12,clark12,firn12}, due to a lack of observational data, the nature of Bsgs is a long-standing issue in stellar evolution. \cite{hunter08} argued on the basis of high-resolution VLT-FLAMES data that the slowly rotating Bsgs in the Large and Small Magellanic Clouds (LMC; SMC) are post-MS objects (although their large numbers would remain unexpected). \cite{vink08} and \cite{vink10} established a general lack of fast rotating Bsgs ($v$sin$i>50$ km\ s$^{-1}$) and suggested that these objects could naturally be explained as MS stars, if they lose their angular momentum via an increased mass-loss rate due to the bi-stability jump (the so-called bi-stability braking). They also pointed out that this mechanism would be efficient if the stars spend significant amounts of time on the MS. While the first hypothesis received some support by the apparently brightest supernova (SN) in the telescopic era: SN\,1987A, bi-stability braking for stars with initial masses above 40\,$M_\odot$ was confirmed by \cite{markova14}.

  Hot stars with high mass-loss rates are expected to have strong emission lines, predominantly formed by recombination. The H$\alpha$ line is an excellent tracer for hot-star mass loss, as its detection is possible for large numbers of stars. Moreover, the strong velocity dependence of H$\alpha$ combined with high-resolution spectroscopy could provide valuable information about velocity fields and density structures in Luminous Blue Variables (LBVs) and SN progenitors \citep{groh11}. That is why understanding the behaviour of the H$\alpha$ line is crucial to our understanding of mass loss.

  After this introduction, in Sect.\,2, we describe shortly the used input parameters for our supergiant models. In Sect.\,3, we show that the computed line profiles exhibit non-monotonic temperature behaviour, and provide the most likely explanation for it. Section 4 explores the impact of clumping on H$\alpha$ line equivalent width (EW) and optical depth. In Sect.\,5 we discuss the importance of our findings and finally in Sect.\,6 we summarise our results and discuss  possible implications for future work.

\vspace{\vfac}
\section{Method and input model parameters}
\label{CMFGEN}

  In order to study H$\alpha$ line formation, we use the non-LTE radiative transfer code {\sc cmfgen} \citep{hillier98,hillier01}, and we calculate a model grid  over a range of temperature and log\,g appropriate for Bsgs (cf. Table~\ref{table:tab1}). Apart from changes in $T_{\rm eff}$ and log\,g, H$\alpha$ line formation is sensitive to luminosity, $L_\star$, mass, $M_\star$, mass-loss rate $\dot{M}$, clumping, and velocity structure. As {\sc cmfgen} does not solve the momentum equation of the wind, a velocity structure needs to be assumed. For the accelerating part of the wind, we adopt a standard $\beta$-type velocity law with $\beta=1$, whilst a hydrostatic solution is adopted for the subsonic region.

  To first understand how the H$\alpha$ line profile changes in a {\it qualitative} sense, we use models with a simplistic H chemical composition, keeping $L_\star$, $M_\star$, $\varv_{\infty}/\varv_{\rm esc}$ and $\dot{M}$ fixed, whilst we vary $T_{\rm eff}$ over a range from 30\,000 K down to 12\,500 K. Following \cite{vink01}, we adopt the parameters listed in Table~\ref{table:tab1}. 
 
  Note that for fixed $L_\star$, the changes in  $T_{\rm eff}$ inevitably lead to different $R_\star$ and $\varv_{\rm esc}$ values, which may influence the H$\alpha$ EW (to account for this, we used an optical depth parameter tool that was introduced by \cite{puls96}; see Sect.~3.1). To keep as many model parameters as possible fixed, we used $\varv_{\infty}/\varv_{\rm esc}=2$ -- the mean value of $\varv_{\infty}/\varv_{\rm esc}$ ratio at both sides of the bi-stability jump \citep{lam95,markova08}. It is more natural to keep the $\varv_{\infty}/\varv_{\rm esc}$ ratio constant rather than the value of $\varv_{\infty}$ itself, as the models have different radii.

  It is important to keep in mind that due to metal line blocking and a blend with He\,{\small II} the H$\alpha$ line could be sensitive to changes in He as well as metal abundance. To estimate these effects we also compare how the H$\alpha$ EW behaves in models with different chemical complexities: (i) pure H, (ii) H $+$ He, with two different He mass fractions (Y$_{\rm He}=$ 0.25 and 0.6), as well as (iii) metal line-blanketed models. 

  \begin{table}
    \caption{Adopted stellar parameters used in the model grid.} 
    \centering
    \begin{tabular}{c c c c c  } 
      \multicolumn{5}{c}{log\,$\left(\frac{L_\star}{L_\odot}\right)=5.5$, $M_\star=40\,M_{\odot}$,  $\dot{M}=2.33\times10^{-6} [M_\odot \rm yr^{-1}]$} \\ 
      \hline\hline\noalign{\smallskip}
      $T_{\rm eff}$ [K]   & $R_\star$ [$R_\odot$] & $\varv_{\infty}=2\times \varv_{\rm esc} [\frac{\rm km}{\rm s}]$ & log(g)     & model\\ 
      \hline\noalign{\smallskip}
      30\,000	  & \ 21 & 1701   &	 3.40  \\
      27\,500	  & \ 25 & 1558   &	 3.25  \\ 
      25\,000	  & \ 30 & 1447   &	 3.09  \\ 
      22\,500	  & \ 37 & 1276   &	 2.90  & M \\ 
      20\,000	  & \ 46 & 1138   &	 2.70  \\ 
      17\,500	  & \ 61 & \ 998  &	 2.47  \\ 
      15\,000	  & \ 83 & \ 855  &	 2.20  \\ 
      12\,500	  &  120 & \ 715  &	 1.88  & C \\ 
    \end{tabular}
    \label{table:tab1} 
  \end{table}

\vspace{\vfac}
\section{Results}
  \label{Results}
  \subsection{H$\alpha$ line profile and equivalent width}

  Our systematic examination of the H$\alpha$ line for supergiant models over $T_{\rm eff}$ range between 30\,000 and 12\,500\,K shows non-monotonic changes in the H$\alpha$ line profiles with $T_{\rm eff}$. As shown in Fig.~\ref{fig:HaLP}, the H$\alpha$ line becomes stronger when $T_{\rm eff}$ drops from 30\,000 to 22\,500\,K, where the  line reaches its peak value. Below 22\,500\,K the line becomes weaker when the effective temperature is further reduced to 12\,500\,K.    

  Although we have kept all stellar parameters fixed, the models have different radii and $\varv_\infty$. Note that in the hottest models the radii are up to a factor of 6 smaller, whilst the     terminal velocities are only up to a factor 2.4 higher. To extract the {\it true} temperature effect,  the different radii and terminal velocities should be taken into account. One way to do this is through the use of the wind-strength parameter $Q$ concept introduced by \cite{puls96}. It was demonstrated that for O-type stars recombination lines remain unchanged by specific sets of individual values of $\dot{M},v_{\infty}$ and $R_\star$ as long as the wind-strength parameter \textit{Q}:
    \begin{equation}
      Q = \frac{\dot{M}}{(v_\infty R_\star)^{1.5}}
    \label{eq_Q}
    \end{equation}
  is invariant and $T\rm _{eff}$ remains unchanged. Here $\dot{M}$ is calculated in $M_\odot/$yr, $v_\infty$ in kms$^{-1}$ and $R_\star$ in $R_\odot$. As an example, in Fig.~\ref{fig:HaLP} we show that the H$\alpha$ line profile from model C (red dashed line) is basically unaffected when the mass-loss rate is decreased by factor of two and the radius by $2^{2/3}$ (black triangles).

  The right-hand panel of Fig.~\ref{fig:HaLP} displays how the H$\alpha$ EW behaves as a function of $T\rm _{eff}$  for models with similar  parameters, but with different individual $\dot{M}$ which scale in such a way that $Q$ is constant (blue squares). The peak is still present and this implies that the H$\alpha$ behaviour is not due to the changes in $v_\infty\times R_\star$, i.e., it is a real temperature effect. Note that the peak is slightly shifted towards $T_{\rm eff}$ $\sim$ 20\,000\,K.

\begin{figure} \centering
    \resizebox{0.9\hsize}{!}{\includegraphics{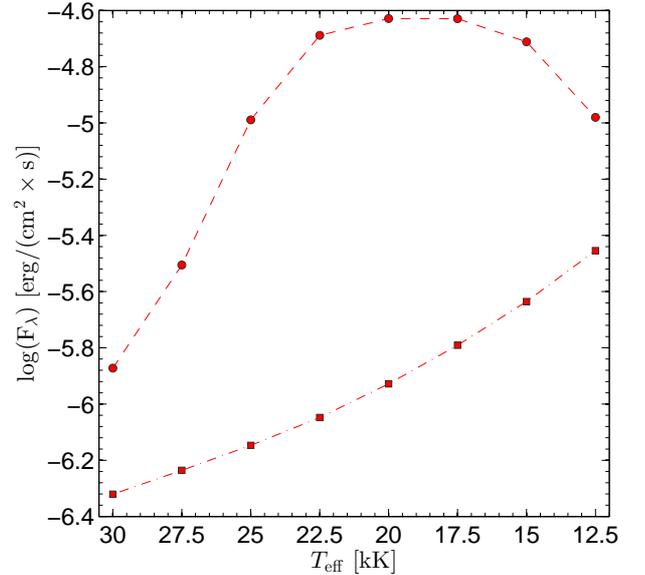}}
    \caption{Integrated line (circles) and continuum flux (squares) at the wavelength of H$\alpha$ for H $+$ He models. Note that the flux represents the flux at a distance of 10 parsec.}
    \label{fig:lflux}
  \end{figure}

  To gain additional insight, the EW was separated into the line and continuum flux (respectively illustrated in Fig.~\ref{fig:lflux}  with red circles and   squares). The reason for doing so is that the EW is a measurement for the ratio of the line flux over the continuum flux, and models with similar EW may have completely different line fluxes. As can be seen in  Fig.~\ref{fig:lflux}, the coolest model  has a seven times larger line flux than the line flux in the hottest model, whilst in Fig.~\ref{fig:HaLP} it seems that both models have a similar line strength. The similar H$\alpha$ EW in both models is due to a constantly increasing continuum flux when reducing $T_{\rm eff}$, and the higher line flux in the coolest model. The peak in right-hand panel of Fig.~\ref{fig:HaLP} results  at 22\,500\,K  (and not at 20\,000\,K as in Fig.~\ref{fig:lflux}) because the ratio between the line and continuum fluxes is the largest for that $T_{\rm eff}$. Note  that the cooler supergiants models have larger radii which increases the continuum flux with decreasing $T_{\rm eff}$.

  The right-hand panel of Fig.~\ref{fig:HaLP} also presents behaviour patterns in models with different chemical complexities. The simplest set of models (with only H; i.e. without He) presented       with red crosses show practically identical H$\alpha$ EWs\footnote{We have defined the line EW to 	be positive for an emission line and negative for an absorption line.} as the EWs from models including both H and He (displayed with grey circles). In other words, He does not seem to influence the H$\alpha$ line. The reason could be that for this $T_{\rm eff}$ range He\begin{small}II\end{small} is diminished and the blueward He\begin{small}II\end{small} blend is not essential, whilst the HeI continuum plays only a minor role at these temperatures.       However, the latter holds only when the He abundance is less than 25\% by mass. The red asterisks in the right-hand panel demonstrate that the He rich models (comprising 60\% He mass fraction) have a factor 2-3 lower H$\alpha$ EW. This       behaviour is in agreement with the results of \cite{dimitrov87} who found that a high abundance of He produces stronger absorptions in H lines. Despite the quantitative differences, He is not important for the {\it qualitative} behaviour of H$\alpha$ versus $T_{\rm eff}$. 

  The line EW values from the simplistic H+He models (grey circles) are also compared to those determined from more realistic chemical models (green triangles), which include C, N, O, Si, S, P and Fe (with atomic data as listed in the table from  Appendix B and half-solar metal abundances). Whilst there are quantitative differences due to line blanketing, the {\it qualitative} behaviour in H$\alpha$ EW is similar in both sets of models with different complexities. This implies that the reason for their behaviour is fundamentally driven by the properties of H. Despite their simplification, the models including H+He only (with atomic data as indicated in  Table~\ref{table:tab2}), provide an overall picture of the effective temperature dependence of H$\alpha$ for Bsgs. It is therefore reasonable to take advantage of these H+He models, using them as a starting point for our investigation.    

  As an aside, we found that for models with a flatter velocity law ($\beta=2$) or with different mass-loss rates (varying from $\sim 10^{-7}$ to $10^{-5}$ $M_{\odot}$yr$^{-1}$) the resulting H$\alpha$ EW behaviour was qualitatively similar to those presented in Fig.~\ref{fig:HaLP} with H$\alpha$ EW showing a peak at $T_{\rm eff}$ $\simeq$ 22\,500\,K.
  \vspace{\vfac}
  \begin{table}
    \caption{Atomic data used for our simplistic H $+$ He supergiant models. For each ion the number of full levels, super levels, and bound-bound transitions are provided.}
    \centering
    \begin{tabular}{l r r r } 
      \hline  \hline 
      Ion & Super levels   & Full levels & b-b transitions \\
      \hline
      H  I      &   20 &  30  &   435   \\ 
      He I      &   45 &  69  &   905   \\ 
      He II     &   22 &  30  &   435   \\ 
    \end{tabular}
    \label{table:tab2}
  \end{table}

  \begin{figure}
    \centering
    \resizebox{\factor\hsize}{!}{\includegraphics{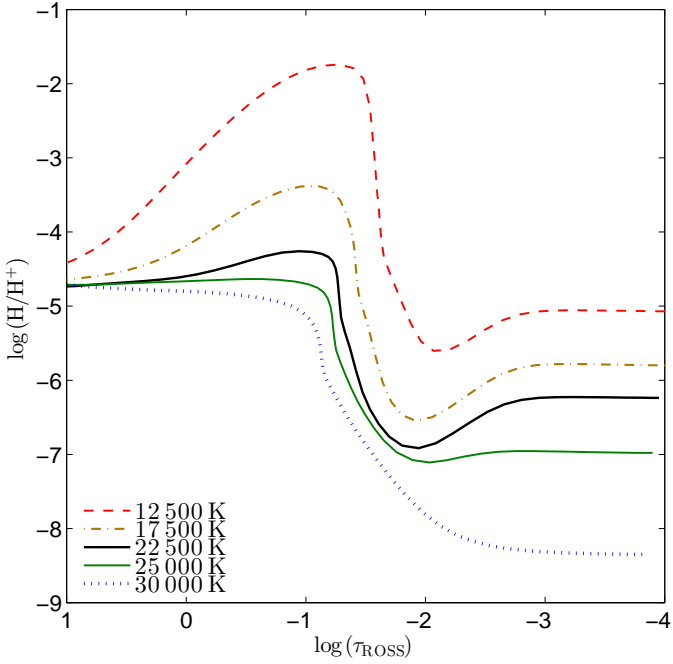}}
    \caption{Hydrogen ionisation structure for models with various $T_{\rm eff}$.}
    \label{fig:H_ION}
  \end{figure}

  \begin{figure}
    \centering
    \resizebox{\factor\hsize}{!}{\includegraphics{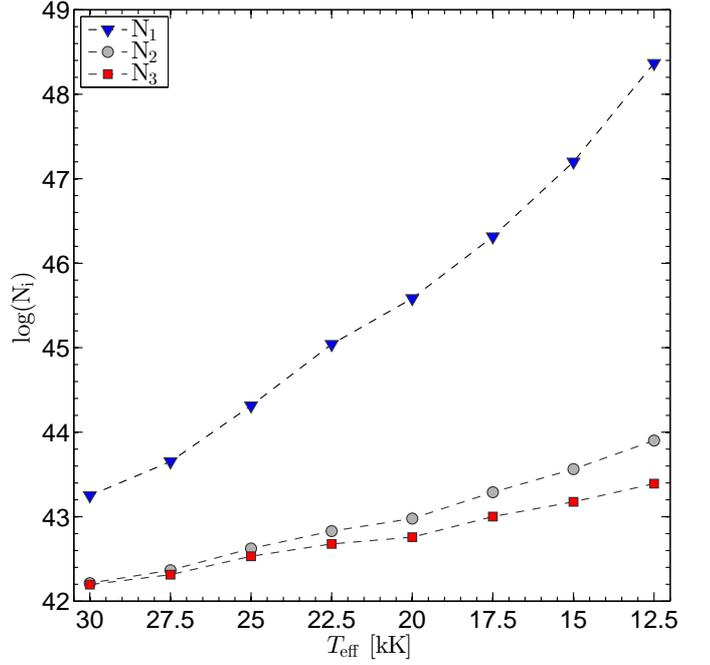}}
    \caption{Total number of H atoms in the stellar wind versus $T_{\rm eff}$. Note that the total number of H atoms is determined from $\rm \tau_{ross}<2/3$.}
    \label{fig:totNi}
  \end{figure}
  \vspace{\vfac}

  \subsection{Two branches of H$\alpha$ behaviour}
    \label{two_branches}

    \begin{figure}
      \centering \resizebox{\factor\hsize}{!}{\includegraphics{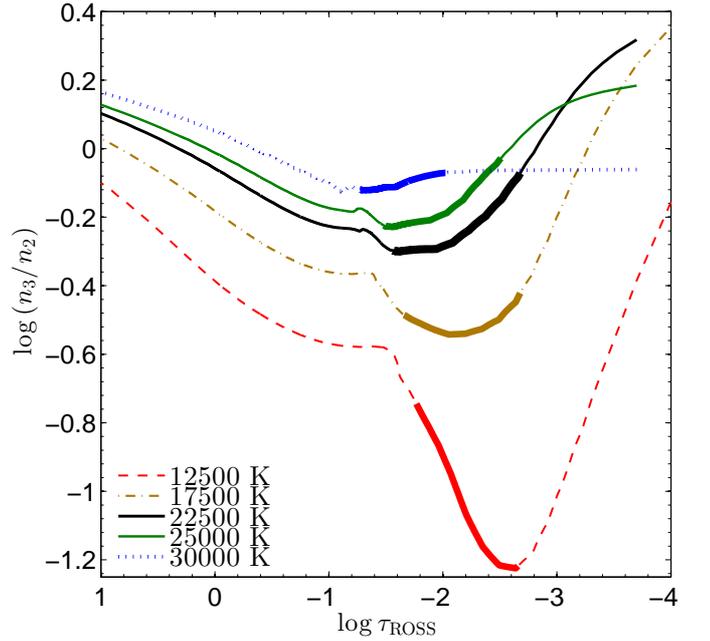}}
      \caption{Changes in the $(n_3/n_2)$ ratio with $T_{\rm eff}$. {\bvp Regions} where most of the emergent H$\alpha$ photons originate from are represented with a {\bvp thick} solid line (cf. Appendix A).}
      \label{fig:H32}
    \end{figure}

    \begin{figure}
      \centering
      \resizebox{\factor\hsize}{!}{\includegraphics{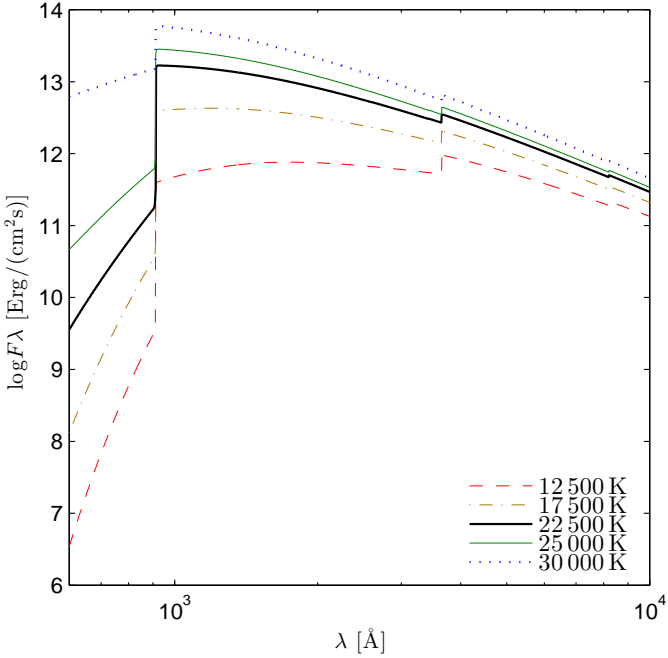}}
      \caption{Spectral energy distribution at the stellar surface of our models.}
      \label{fig:SED}
    \end{figure}

    The H$\alpha$ EW changes drastically with $T_{\rm eff}$ in all sets of models. In general, we find there to be two branches in Figs.~\ref{fig:HaLP} and~\ref{fig:lflux}: the  ``hot'' and ``cool'' ones. The hot branch is located between 30\,000 to 22\,500\,K, where the  H$\alpha$ line emission becomes stronger with decreasing $T_{\rm eff}$. At the cool branch, from 22\,500 to 12\,500\, K, the behaviour of the line flux changes in the opposite direction. This implies that there is a qualitative change in the behaviour of H$\alpha$ around 22\,500\,K. 

  \subsubsection{The ``hot''  branch}

    \begin{figure}
      \centering
      \resizebox{\hsize}{!}{\includegraphics{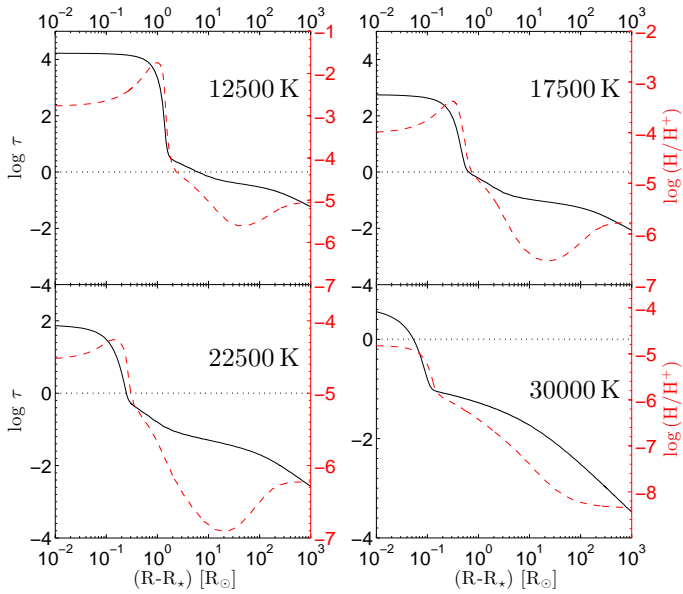}}
      \caption{Comparison between the H ionisation structure (red dashed line) and the Lyman continuum optical depth at $\lambda \sim 900 $ \AA \ (black solid line) versus the distance from the stellar photosphere. {\bvp Solid} lines are reserved for the wind optical depth, whilst the dotted horizontal lines indicate the transition between optically thick and thin part of the wind in the Lyman continuum ($\tau=1$). {\bvp Red} colour on the right-hand side is used for the H ionisation structure.}
      \label{fig:TAUcon}
    \end{figure}

    The formation of the H$\alpha$ line involves transitions between the 3$\rm ^{rd}$ and 2$\rm ^{nd}$ level of H. Therefore, the emission should be proportional to the {\bvp total number of H atoms in the 3$\rm^{rd}$ level  in the wind above $\tau_{\rm ROSS}=2/3$ ($N_3$)}. As $N_3$ (as do the number of H atoms in other levels) scales with the fraction of neutral H, we show the H ionisation structure of our models in  Fig.~\ref{fig:H_ION}. It is evident that the wind is mostly ionised in all models and as $T_{\rm eff}$ drops the wind recombines slightly. The fraction of neutral H increases to almost two percent when $T_{\rm eff}$ is reduced to 12\,500\,K (at log\,$\tau_{\rm ross}$ $\sim$ $-$1.5, log\,(H/H$^{+}$) $\sim$ -1.8). Furthermore, in an absolute sense the total number of H atoms in the second level ($N_2$) and $N_3$ increase, as illustrated in Fig.~\ref{fig:totNi}. As a result, the flux in H$\alpha$, which is proportional to $N_3$, should to first order increase as $T_{\rm eff}$ drops. Therefore, the trend in H$\alpha$ on the ``hot`` side of  Figs.~\ref{fig:HaLP} and  ~\ref{fig:lflux} can be understood in terms of a simple recombination effect (see also Puls et al. 1996). Difficulties with such a simple explanation would arise if we were to try to explain the existence of the cool branch in a similar way. Figure~\ref{fig:totNi} illustrates that when $T_{\rm eff}$ is reduced from 22\,500 to 12\,500\,K, $N_3$ is still increasing and this should probably produce a stronger H$\alpha$ line for cooler models. Contrary to expectation, the opposite behaviour of H$\alpha$ is produced for this branch. 

  \subsubsection{The ``cool''  branch}

    On the cool branch, small changes in $T_{\rm eff}$ lead to qualitatively different H$\alpha$ line profiles. Most notable is the appearance of a P\,Cygni profile in  Fig.~\ref{fig:HaLP} when the effective temperature is reduced from 15\,000 to 12\,500\,K. In fact, these  differences between both H$\alpha$ line profiles contain major clues to the unexpected changes in the line flux of the models on the cool branch. In case H$\alpha$ were a pure recombination line only the third level would be relevant. {\bvp However, as shown later,  H$\alpha$ increases its optical depth at cooler $T_{\rm eff}$.  It is thus  necessary to assess the source function and  the ratio of the number of H atoms per cm$^3$ in the 3$\rm^{rd}$ over the 2$\rm^{nd}$ level ($n_{3}/n_{2}$).}
   
    Figure~\ref{fig:totNi} reveals that $N_3$ is constantly increasing when $T_{\rm eff}$ is reduced, however, $N_2$ increases more steeply. The $(n_3/n_2)$ ratio has been plotted versus Rosseland optical depth in  Fig.~\ref{fig:H32}. First of all, the $(n_3/n_2)$ ratio is close to unity for the  models at the ``hot'' branch.  However, for the models on the cool branch, $n_2$ becomes significantly larger than $n_3$, particularly in the outer wind. Although $N_1, N_2, N_3$ are all increasing as $T_{\rm eff}$ drops, the {\it ratio} $(n_3/n_2)$ is always decreasing, and  the second level  becomes more populated than the third level on the cool branch. This leads to a ``dip'' in $(n_3/n_2)$ in the outer wind of the cooler models. The ``dip''  is quite pronounced, especially for the coolest model, where $n_2$ is ten times higher than $n_3$. The increased $n_2$ in the outer wind produces absorptions for H$\alpha$ photons emitted close to the photosphere. This naturally decreases the H$\alpha$ flux. Hence the question about the decreasing EW over the cool branch, could be referred to as an issue regarding the behaviour of the $(n_3/n_2)$ {\it ratio}.

    Therefore, the next question is why $n_2$ becomes significantly larger than $n_3$? Is the ``dip'' in $(n_3/n_2)$ ratio is predominantly caused by an \textit{increase} of $n_2$ or by a \textit{decrease} of $n_3$? In order to address these questions, we take a detailed look into the behaviour of the continua and their effects on individual levels.

      \begin{figure*}
	\hspace{0.2cm}
	\resizebox{0.95\hsize}{!}
	{\includegraphics{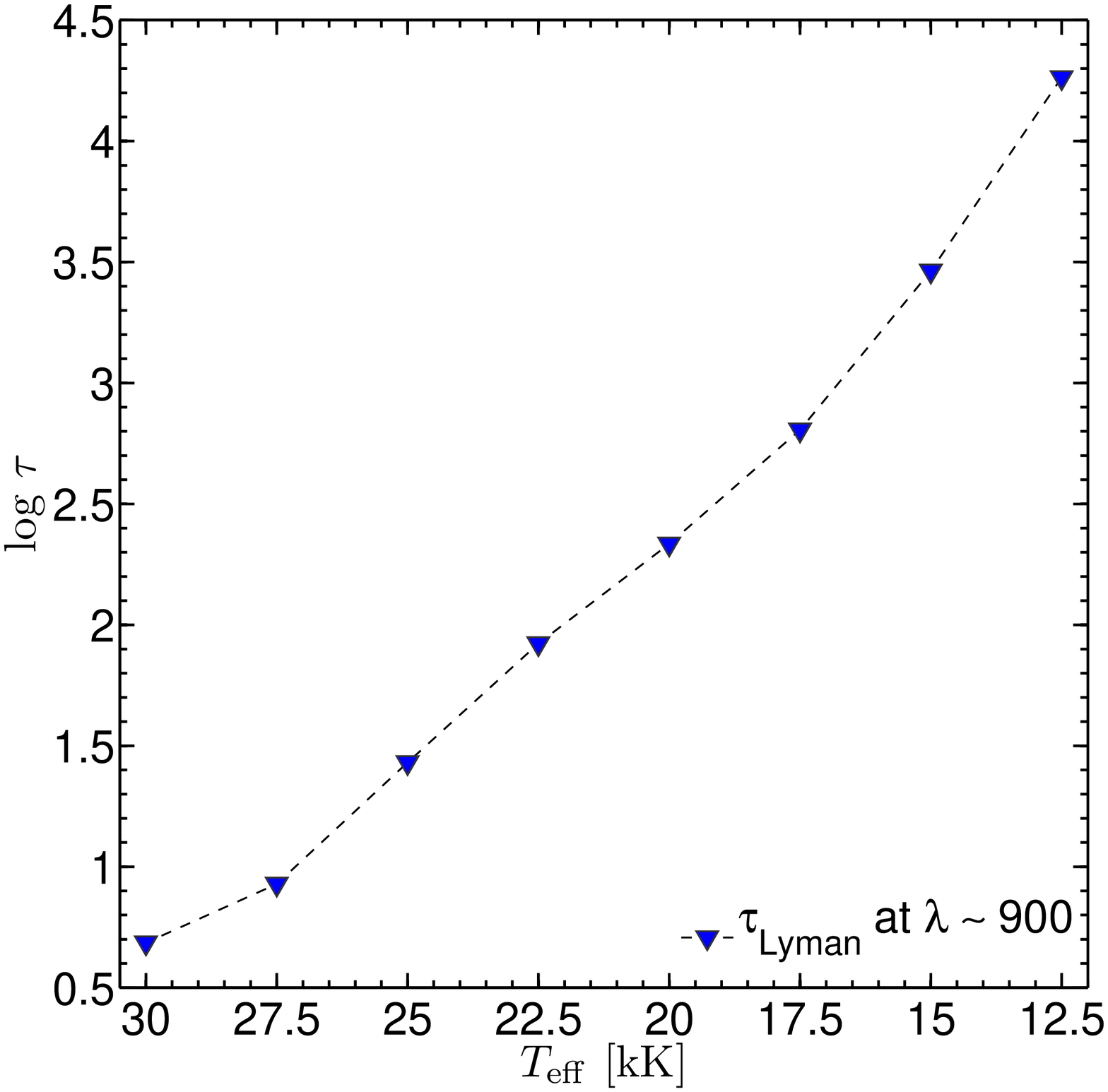}\hspace{\spfactor}\includegraphics{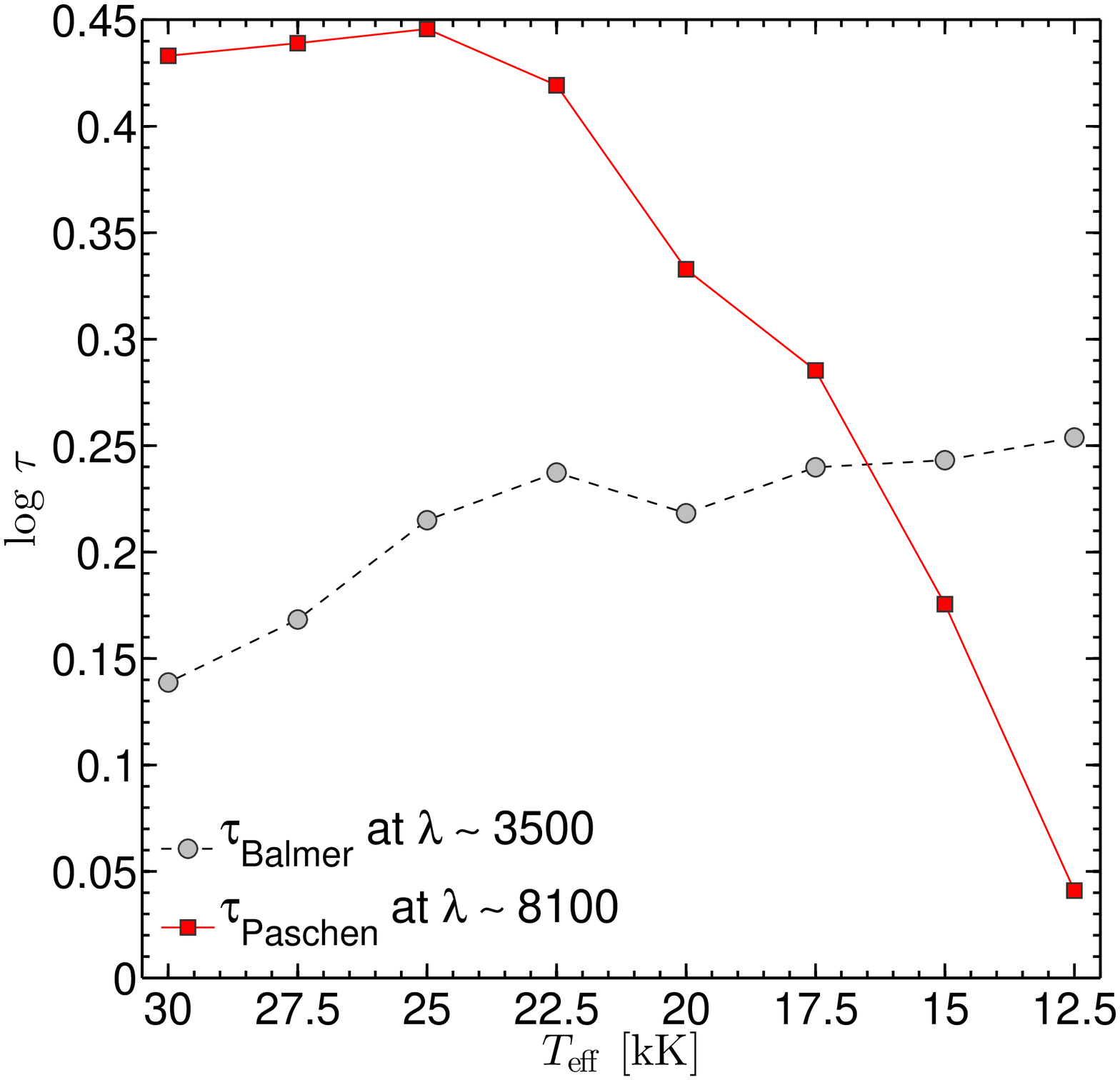}}
	\par
	\caption{Wind optical depth at $\tau_{\rm ross}=2/3$ in the Lyman (left),  Balmer (grey circles) and Paschen continua (red squares) (right).}
	\label{fig:TAU}
    \end{figure*}

  \subsection{Two possible explanations for the existence of the ``cool'' branch}
    \subsubsection{A decrease of $n_3$?}

      In Fig.~\ref{fig:SED} we show spectral energy distributions (SEDs) at the surface. The Lyman continuum flux is greatly reduced from 30\,000 to 12\,500\,K. Despite that, nothing dramatically happens with the Lyman continuum around 22\,500\,K  - the changes in the continuum flux are rather gradual. To understand the behaviour of the Lyman continuum in Fig.~\ref{fig:TAUcon} we  compare the continuum optical depth at $\lambda \sim 900$\,\AA \,(black solid line) to changes in the H ionisation structure (red dashed line with ordinate in red colour placed on the right side).
      It is evident that the Lyman continuum becomes optically thick at distances closer than $d_\tau  \sim $ 0.06, 0.25, 0.6 and 7 [R$_\odot$] from the photosphere (roughly $\tau_{\rm ross} =2/3$), respectively for the models with $T\rm _{eff}$ = 30\,000, 22\,500, 17\,500 and 12\,500\,K.  The fraction of neutral H at those  distances is between $\sim 10^{-4.9} \rm and \ 10^{-4.3}$. The comparison between the Lyman continuum optical depth and H ionisation structure shows that as soon as neutral H atoms exceed \textit{critical} values, the Lyman continuum becomes optically thick. Moreover, the steep increase of neutral H close to the star  leads to a significant increase of  the  optical depth of the Lyman continuum. Consequently, a large fraction of Lyman ionising photons are blocked, and the Lyman continuum is no longer the main source of ionisation.

      \cite{naj97} studied the appearance of H and He lines in the wind of the LBV P\,Cygni, and they found that H recombination, crucial for the Ly continuum and the Ly$\alpha$ optical depth, block the ionising Lyman flux. In their models, the increased $n_2$ is due to the high optical depth of Ly$\alpha$, giving rise to strong absorption, similar to those produced in our coolest B supergiant model.

      As $n_1, n_2$ and $n_3$ are controlled from different continuum ranges, knowledge about the behaviour of  Balmer and Paschen fluxes (not only of the Lyman continuum) is required to understand their behaviour. In Fig.~\ref{fig:SED} it is shown that the Balmer and Paschen continuum fluxes decrease when $T_{\rm eff}$ is reduced. To understand this, we plot the wind optical depth in the Balmer and Paschen continua in Fig.~\ref{fig:TAU} (right panel). Since the cross-section for the photo-ionisation of a H atom in quantum state $n$ by a photon of wavelength $\lambda$ is:      \begin{equation}
	\sigma_{\rm  bf} = 1.31\times 10^{-19}\frac{1}{n^5}\left(\frac{\lambda}{5000 \ \textup{\AA}}\right)^3,
      \end{equation} 
      the opacity would be proportional to $\lambda^3$. Therefore  we have chosen to plot the wind optical depth in the Balmer and Paschen continua at wavelengths close to their corresponding jumps (respectively  $\lambda \sim 3500$ and $8100 \,$\AA).

      \begin{figure*}
	\begin{centering}
	\resizebox{\factor\hsize}{!}{\includegraphics{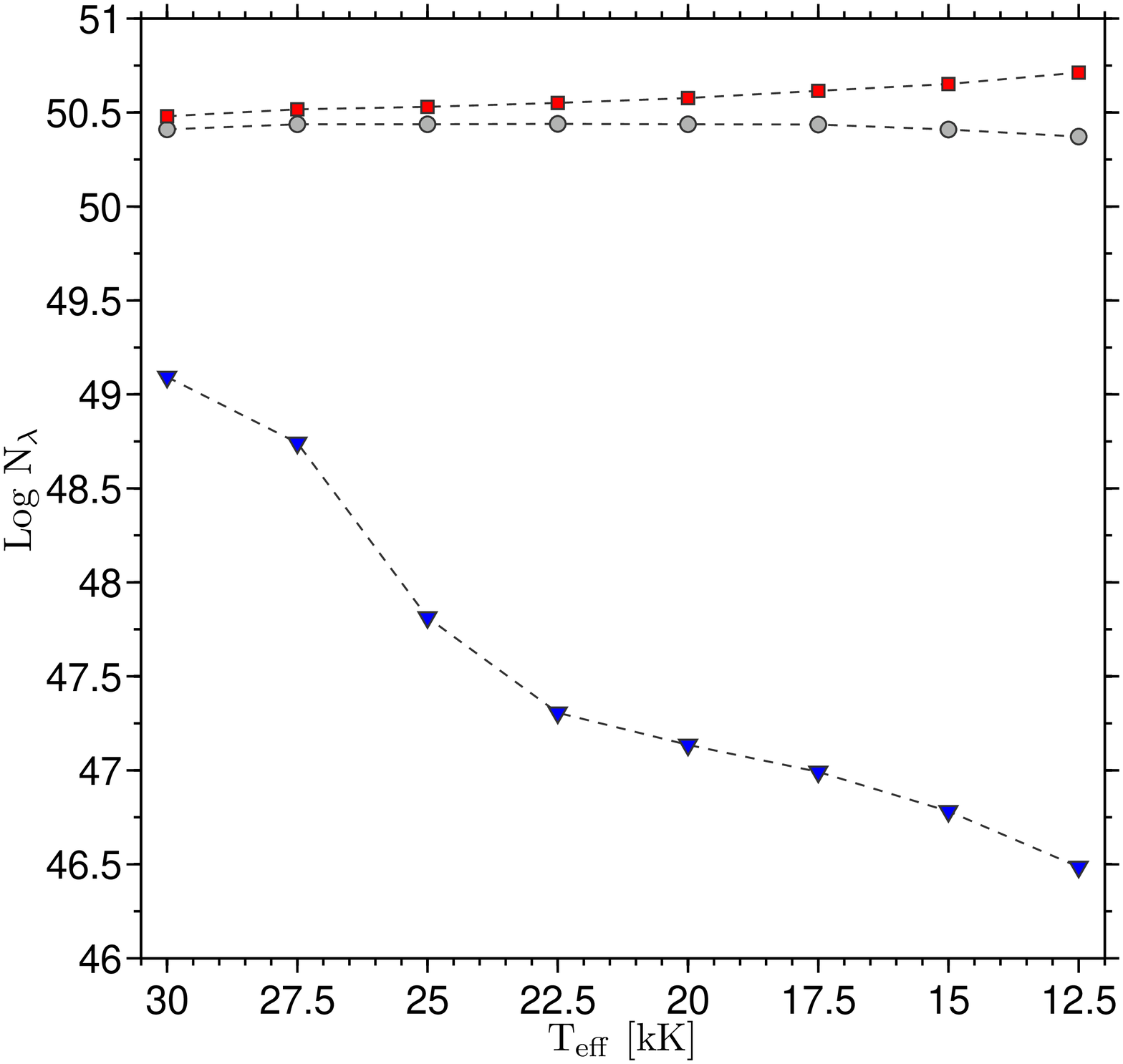}\hspace{\spfactor} \includegraphics{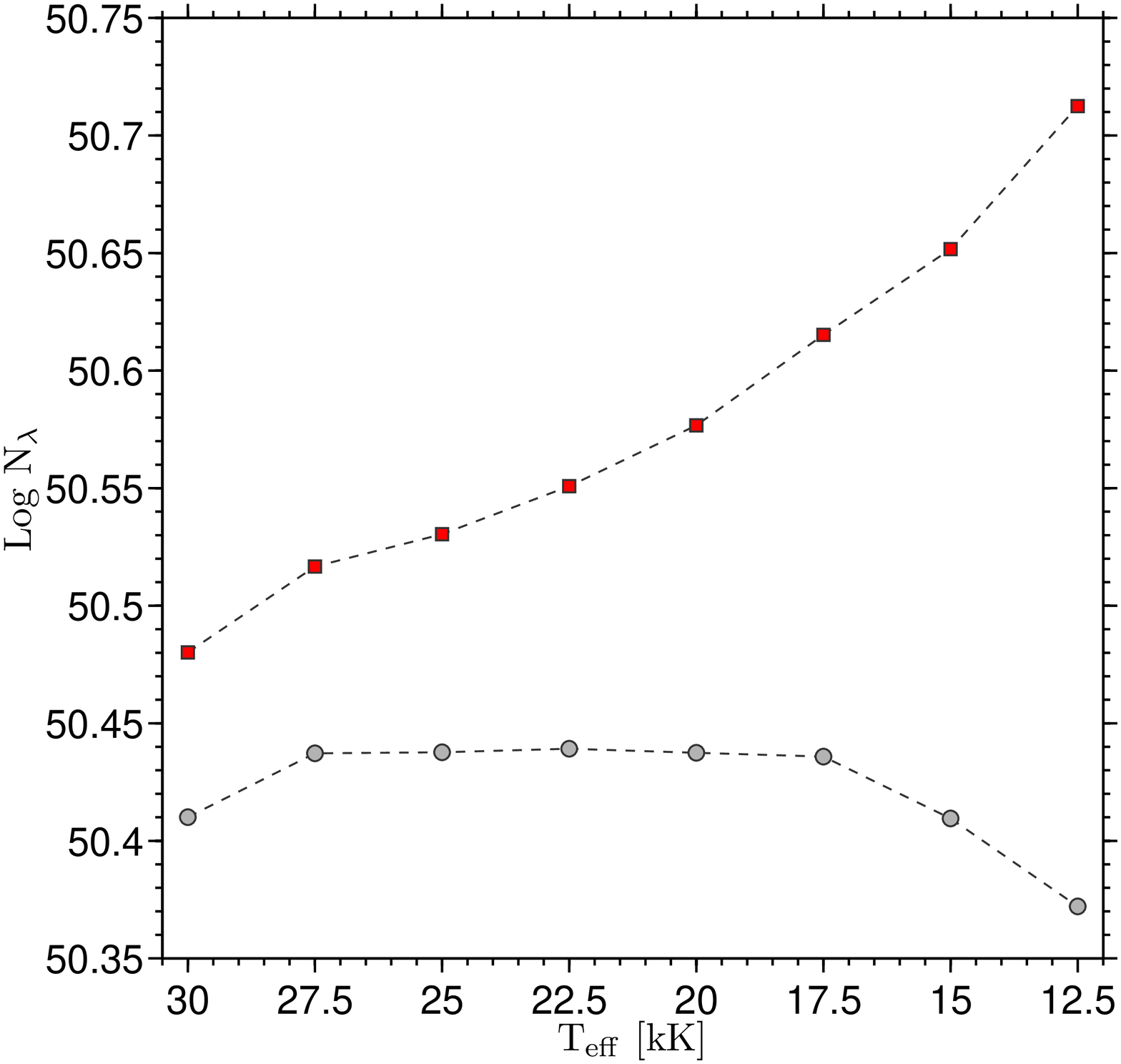}}
	\par\end{centering}
	\caption{{\bvp Number of  photons  in  the Lyman (blue triangles), Balmer (grey circles) and Paschen (red squares) continua vs $T_{\rm eff}$.} {\bvp Right} hand side is a ``zoom in'' from the left hand side.}
	\label{fig:LC}
      \end{figure*}

      Figure~\ref{fig:TAU} shows that the wind optical depth in the Balmer continuum remains fairly constant for all models. Note that when $T_{\rm eff}$ is reduced below 22\,500\,K, the wind optical depth in the Paschen continuum is steeply decreased. Models which are cooler than 17\,500\,K even have $\tau\rm_{Paschen} < \tau\rm_{Balmer} $. This should produce larger changes in the Paschen flux than in the Balmer flux. To quantify the reaction  of the continuum fluxes Fig.~\ref{fig:LC} shows the number of photons - N$_\lambda$, versus $T_{\rm eff}$ for the Lyman, Balmer, and Paschen continua respectively. The number of photons is given by:
      \begin{equation}
	N_\lambda = 4\pi{R_{\star}^{2}}\int_{0}^{\lambda_{c}} \frac{\pi{F_\lambda}\lambda}{hc} \, \mathrm{d} \lambda
      \end{equation}
      where $\lambda_c$ is the wavelength boundary of corresponding continuum series;  $R_{\star}$ is the stellar radius. The computed numbers of photons in the Lyman and Balmer continua are in reasonable agreement (by a factor of $\sim$ 4) with those from previous studies \citep{th84,miller98} if the different radii are taken into account. Balmer and Paschen fluxes are able to ionise H atoms respectively from level 2 and 3 and regulate those levels. It is evident from  Fig.~\ref{fig:LC} (left panel) that these fluxes are about 4 orders of magnitude larger than the Lyman flux at the cool branch. This is a consequence of high wind optical depth in the Lyman continuum (reported in Fig.~\ref{fig:TAU}) and the optically thinner wind in the Balmer and Paschen continua.  Hence, the Balmer and Paschen fluxes are the main sources of ionisation  of H atoms in second and third levels over the cool branch. Figure~\ref{fig:LC} indicates that the  total number of photons able to ionise atoms in level 2 is nearly the same for all models. This would provide nearly the same number of H atoms ionised from the second level over both branches. Therefore, we do not expect dramatic changes in $n_2$ due to the Balmer continuum. By contrast, the ``Paschen photons'' are gradually increasing in number as $T_{\rm eff}$ is reduced. This is seen in the right panel of Fig.\ref{fig:LC}, where we zoomed in around N$_\lambda$ for Balmer and Paschen continua. For cooler models, the increasing flux in the Paschen continuum may thus depopulate more H atoms in the third level.

      To understand the key question whether the ``dip'' in $(n_3/n_2)$ results solely from an increase of level 2, or from a decrease of level 3 as well, we show in Fig.~\ref{fig:ni} how their number densities (and the number density of levels 4 and 10, $n_4$ and $n_{10}$) change with $T_{\rm eff}$. It is evident from the plot that when $T_{\rm eff}$ is reduced, $n_3$ behaves in a fashion more similar to the number densities of the higher levels, e.g. $n_4$. This is expected if the levels are mainly fed by recombinations. Therefore, a decrease of level 3 does not occur and consequently the $(n_3/n_2)$ ratio is not affected by changes in $n_3$. It seems that the changes in the Paschen continuum with $T_{\rm eff}$ are not large enough to cause a decrease of $n_3$. As a result, $n_2$ should play a major role in the $(n_3/n_2)$ ratio, causing the ``dip''.

    \subsubsection{An increase of $n_2$?}

      \begin{figure}
	\centering
	\resizebox{\factor\hsize}{!}{\includegraphics{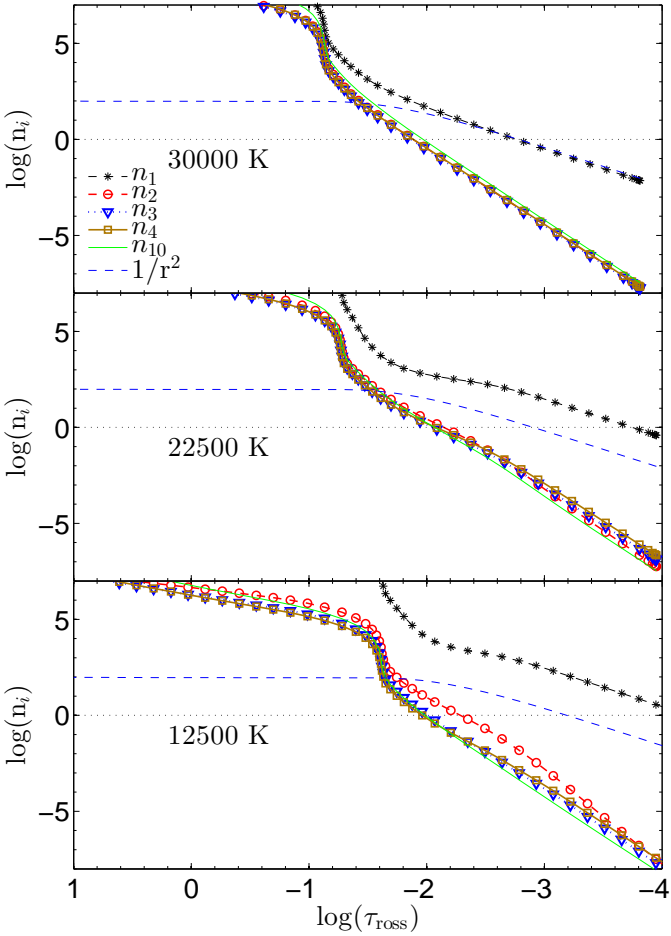}}
	\caption{Population levels of H as a function of Rosseland optical depth.}
	\label{fig:ni}
      \end{figure}

      \begin{figure}
	\centering
	\resizebox{\factor\hsize}{!}{\includegraphics{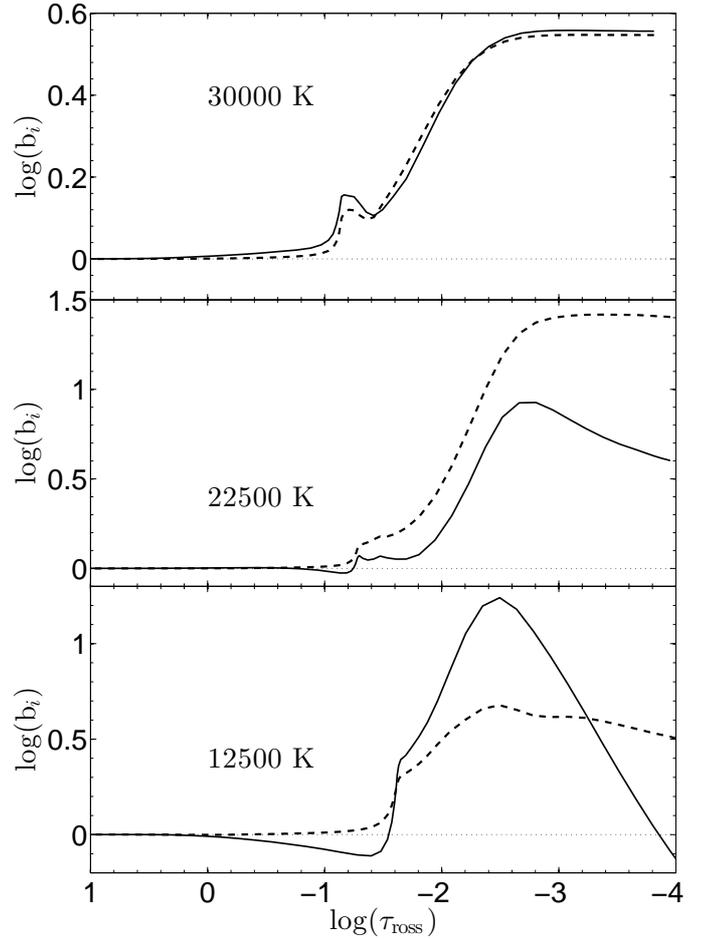}}
	\caption{Non-LTE departure coefficients for the 2$^{\rm nd}$ (solid) and 3$^{\rm rd}$ (dashed) level of H.}
	\label{fig:dc23}
      \end{figure}

      \begin{figure}
        \centering
        {\begin{minipage}{\linewidth}
	  \hspace{0.42cm}\includegraphics[width=0.96\linewidth, height = \textheight, keepaspectratio=true]{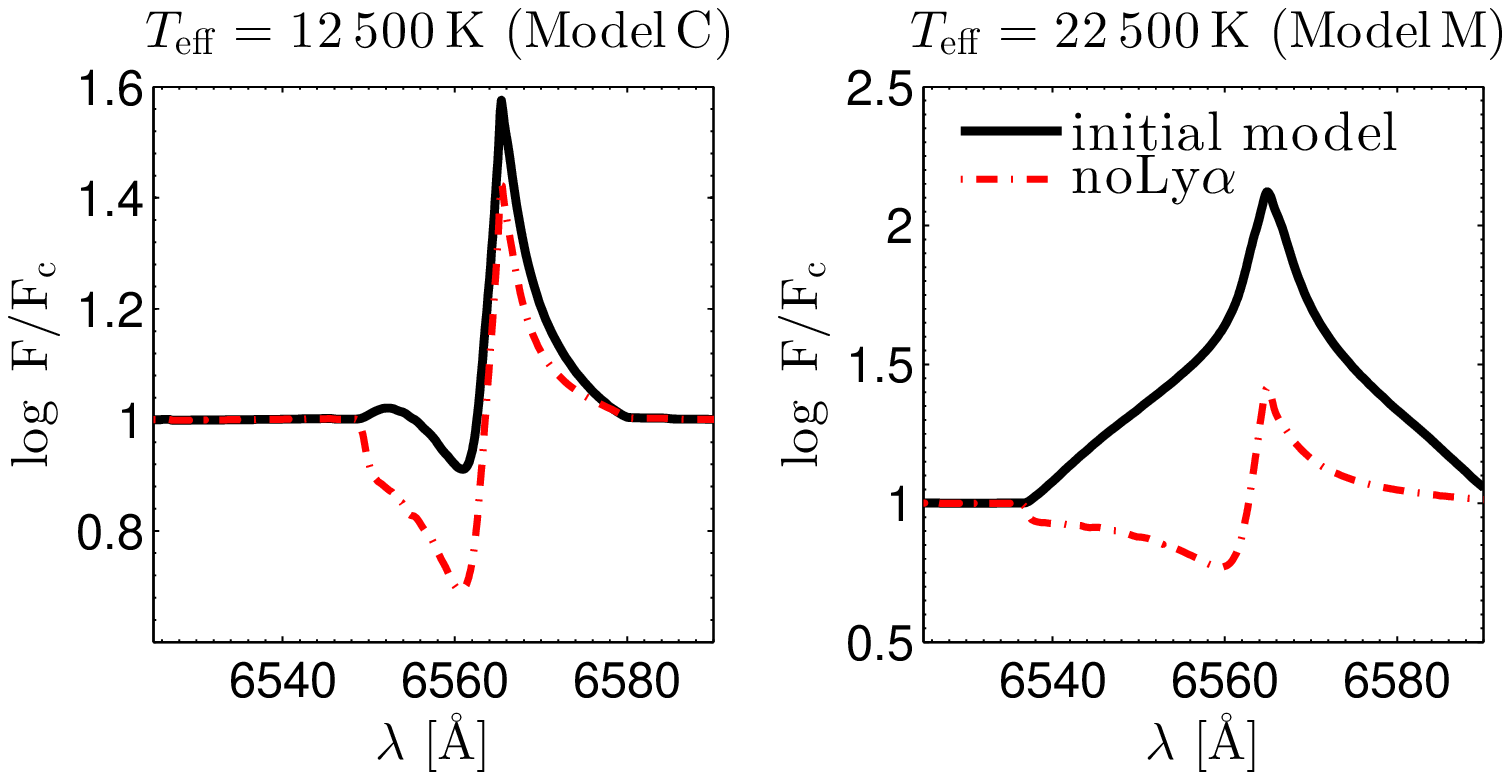}\\
	                
                \includegraphics[width=\linewidth, height = \textheight, keepaspectratio=true]{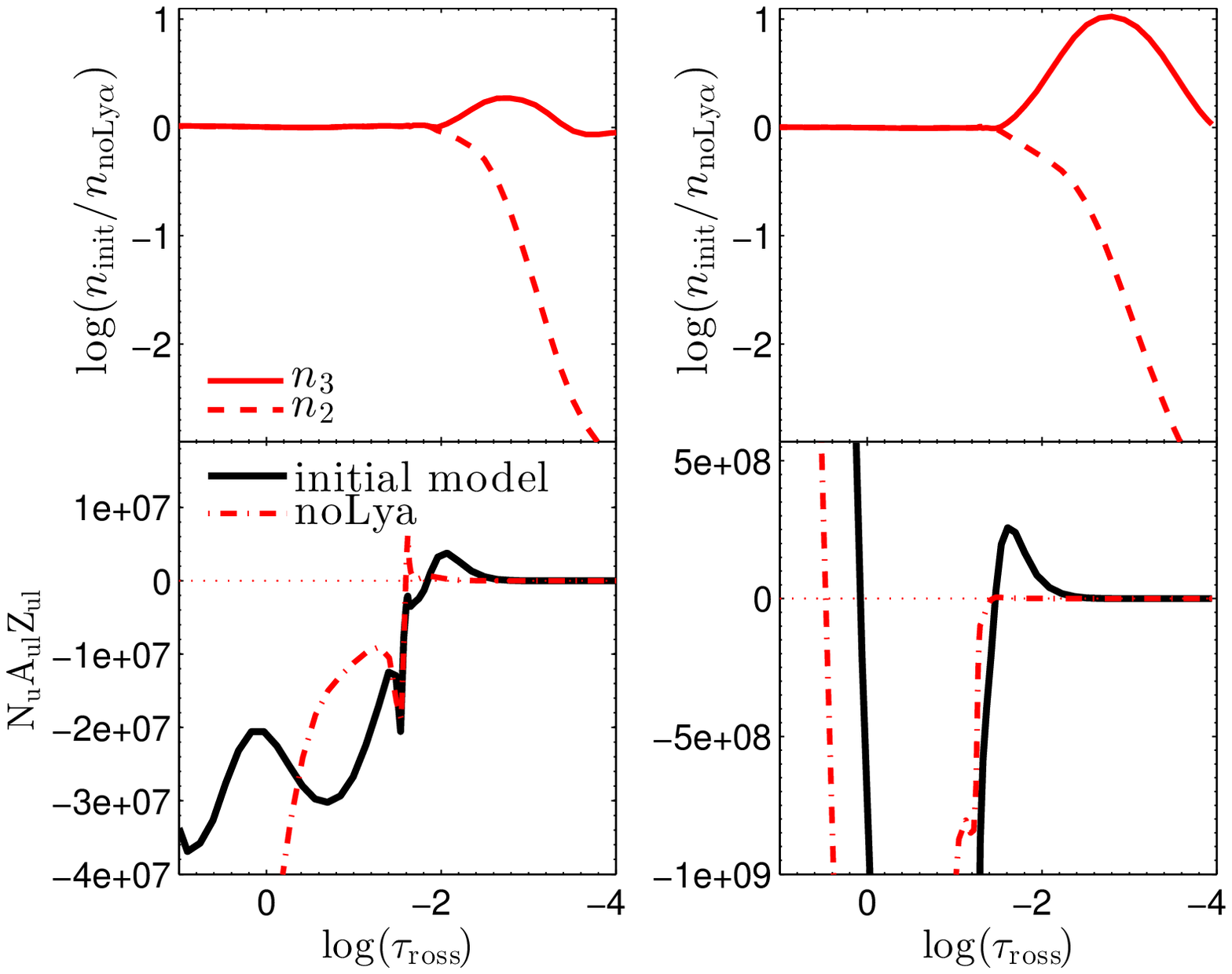}
         \end{minipage}} 
	\caption{{\it Upper panels}: effect of Ly$\alpha$ on the formation of the H$\alpha$ line: initial H$\alpha$ profile (black solid) and the profile from the models in which Ly$\alpha$ transitions were artificially removed (the red dash-dotted line). {\it Middle panels:} changes in the 2$^{\rm nd}$ (dashed) and 3$^{\rm rd}$ (solid) level of H due to the removal of Ly$\alpha$ in model C (left) and M (right). The plots present the ratio of the populations produced from the initial model over the populations from the models without Ly$\alpha$ transition. {\it Lower panels}: comparison of the net radiative rate of 2$\rightarrow$1 transitions in the initial (solid black) and the model without Ly$\alpha$ (the 	red dash-dotted line) (see Sect.~\ref{sub:Lya} for details).}
	\label{fig:TRR_dc}
      \end{figure}

      Due to the enormous optical depth in the Ly continuum (reported in Fig.~\ref{fig:TAU}), the ionising flux is blocked for the ``cool'' models and $n_1$ can no longer be affected by the Lyman continuum. As a result, level 2 becomes the effective ground state, exhibiting a depopulation close to the photosphere, and an overpopulation in the outer wind in model C ($T_{\rm eff}$=12\,500\, K) as shown in Fig.~\ref{fig:dc23}. This is similar to the findings of \cite{puls98} who concluded that $\rm H\alpha$ appears like a P-Cygni line for A supergiants. It should be noted that the produced peak in the departures from LTE for the second level ($b_2$) in model C, occupies the same position as  the ``dip'' in the  $(n_3/n_2)$ ratio. This coincidence indicates that  the ``dip'' is mainly produced by an increase of $n_2$. 

      The difference in behaviour of level 2 compared to higher levels can clearly be seen in Fig.~\ref{fig:ni}, where the number densities of level 3 (blue triangles), 4 (squares) and 10 (green solid line) are similar for all $T_{\rm eff}$. For the models on the ``hot'' branch even $n_2$ behaves in an identical manner to the number densities of the higher levels. This behaviour is to be expected when the level populations are solely fed by recombinations, and when they should scale as $\rho^2\sim r^{-4}$ (via the continuity equation). The population of the ground state is, however, affected by ionisations as well, and it is thus inversely proportional to the dilution factor of the radiation field. The latter effect makes it increase as distance-squared, and in the outer wind the final dependence is 1/r$^2$, as shown in Fig.~\ref{fig:ni}. The black asterisks, connected by a dashed line, represent the number density of the ground state ($n_1$), which is decreasing as 1/r$^2$ (illustrated as a blue dashed line). When $T_{\rm eff}$ is reduced to 12\,500\,K, $n_2$ diverges from higher-lying energy levels and behaves similar to the ground level (decreasing as 1/r$^2$). At the same time the H$\alpha$ line profile forms an absorption component.

    \subsection{Ly$\alpha$ and the second level}
      \label{sub:Lya}

      In order to understand why level 2 diverges from the higher levels when $T_{\rm eff}$ drops, we ran additional models in which we artificially removed the Ly$\alpha$ transition by reducing its oscillator strength by a huge factor ($10^4$). The reaction of H$\alpha$ is displayed in the upper panels of Figure~\ref{fig:TRR_dc}. It highlights the key role of Ly$\alpha$. The effects on model M ($T_{\rm eff}$=22\,500\, K) are striking: H$\alpha$ now switches from a pure emission line into a P\,Cygni line, and the line flux is decreased significantly. Furthermore, for the cooler model the removal of Ly$\alpha$ leads to a deeper absorption component. In the middle panels in Fig.~\ref{fig:TRR_dc} we present produced changes in $n_2$ (dashed) and $n_3$ (solid) due to the absence of Ly$\alpha$ -- as a function of $\rm \tau_{ross}$.  The middle panels demonstrate that the Ly$\alpha$ removal leads to a tremendous increase of $n_2$ (in comparison to the initial model) in the outer wind, leading to a stronger absorption component in  model C, and the appearance of a P\,Cygni profile in model M, where it is noted that the lack of Ly$\alpha$ leads to significant changes in the third level as well: $n_3$ is surprisingly reduced, and as a result the H$\alpha$ flux decreases.

      To understand the changes in the level populations due the removal of Ly$\alpha$, we show its net radiative rate in the lower panels of Fig.~\ref{fig:TRR_dc}.  This quantity is defined as the difference between the number of radiative transitions from the upper (second) level to the lower (ground) state and the number of radiative transitions from the ground level to the second level. As a result, the net radiative rate is positive when there is a net decay of electrons from the upper level, and it is negative when there is a net excitation of electrons. It is seen that as soon as $n\rm _{2noLy\alpha}$ and $n\rm _{3noLy\alpha}$ diverge from their initial ones (at $\rm log(\tau_{ross})<\sim-1.8$ (in model C) and $\rm log(\tau_{ross})<\sim -1.5$ (in model M)) the total Ly$\alpha$ radiative rate in the initial model (black solid line) is positive\footnote{Note that at $\rm log(\tau_{ross})<\sim-1.8$ and $\rm log(\tau_{ross})<\sim-1.5$ the H$\alpha$ line starts to
      form in models C and M (see Appendix\,A for details about the H$\alpha$ forming regions presented in Fig.~\ref{fig:H32}).}. Therefore, the line effectively acts as a ``drain'' for the second level. When we remove it, the decay of electrons from the second level is suppressed, and $n_2$ is tremendously increased, as shown in the middle panels.

      In other words, by artificially removing Ly$\alpha$, we can simulate the appearance of P\,Cygni profiles for hotter models, showing that the Ly$\alpha$ line is key to the H$\alpha$ behaviour. We also note that neither He nor Fe are directly required for achieving this, i.e. it is a pure H effect (as was shown in  Figure~\ref{fig:HaLP}).

\begin{figure}[t]
      \centering
      \resizebox{\factor\hsize}{!}{\includegraphics{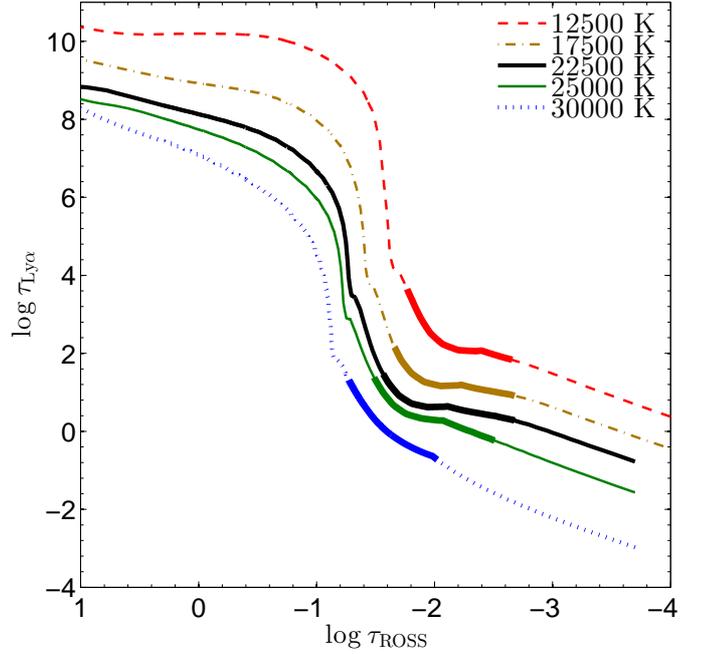}}
      \caption{Lyman $\alpha$ Sobolev optical depth as a function of $\rm \tau_{ross}$. The region where most of the emergent H$\alpha$ photons originate from are shown with  {\bvp thick} solid lines (cf. Appendix\,A)}.
      \label{fig:TAULY}
    \end{figure}

      Figure~\ref{fig:TAULY} illustrates how the Ly$\alpha$ optical depth changes. The location where most of the H$\alpha$ photons  originates form (Appendix A) is shown with thicker line sections. It is evident that at the hot branch Ly$\alpha$ is optically thick in the inner H$\alpha$ forming region and becomes optically thin in the outer region. Furthermore, the Ly$\alpha$ optical depth at
      the start of the H$\alpha$ line-formation region is similar  for hot models. However, when $T_{\rm eff}$ drops below 22\,500\,K, the Ly$\alpha$ optical depth at the start of the H$\alpha$ line-formation region steeply increases, which continues over the cool branch where Ly$\alpha$ is always optically thick throughout the entire H$\alpha$ line-forming region. This means that most photons from the photosphere at the Ly$\alpha$ wavelength do not manage to escape. I.e., they are being scattered or absorbed. As a result, the decay from the second level is effectively shut off. This       can be seen in the lower panels of Fig.~\ref{fig:TRR_dc}.  In  the region where $n_2$ and $n_3$ are affected by Ly$\alpha$ (i.e.\ for $\rm log(\tau_{ross})<\sim-1.8$ for the cool model and $\rm log(\tau_{ross})<\sim -1.5$ for the hotter model), the net radiative rate is 2 orders of magnitude lower for the cooler model, i.e., the second level is depopulated less efficiently by    $2\rightarrow1$ transitions.
 
      In short, the departure of $n_2$ from the higher level occupation numbers (when $T_{\rm eff}$ drops from 22\,500\,K downwards) is related to the Ly$\alpha$ optical depth. When $T_{\rm eff}$ is reduced below 22\,500\,K, the Ly$\alpha$  optical depth	increases steeply, and level 2 is less efficiently depleted through decay. Unlike the second level, level 3 behaves like higher levels, thus the ``dip'' in $(n_3/n_2)$ is the result of an  	increased level 2 population.

  \section{The effect of clumping}

    Currently, {\sc cmfgen} takes optically thin (micro) clumping into account, i.e.,     the clumps are assumed to have a dimension smaller than the photon      mean free path. The density $\rho$ within clumps is assumed to      be enhanced by a clumping factor $D$ compared to the      wind mean density $\bar{\rho}$. This factor can also be understood in	terms of volume filling factor $f=D^{-1}$, assuming that the	inter-clump medium is void.    Mass-loss diagnostic techniques that are linearly dependent on density are insensitive to clumping, whilst recombination lines (sensitive to $\rho^2$) tend to overestimate the mass-loss rate of a clumped wind by a factor of $\sqrt{f^{-1}}$.

    However, if the clumps are optically thick       (macro-clumping), photons may leak through gaps between clumps,      and the mean opacity is affected by the distribution, the      geometrical size, and the shape of the clumps.    In this case the micro-clumping approach is no longer justified.

    \subsection{The H$\alpha$ line  in a micro-clumping approach}
    \begin{figure*}
      \centering
      \resizebox{\factor\hsize}{!}{\includegraphics{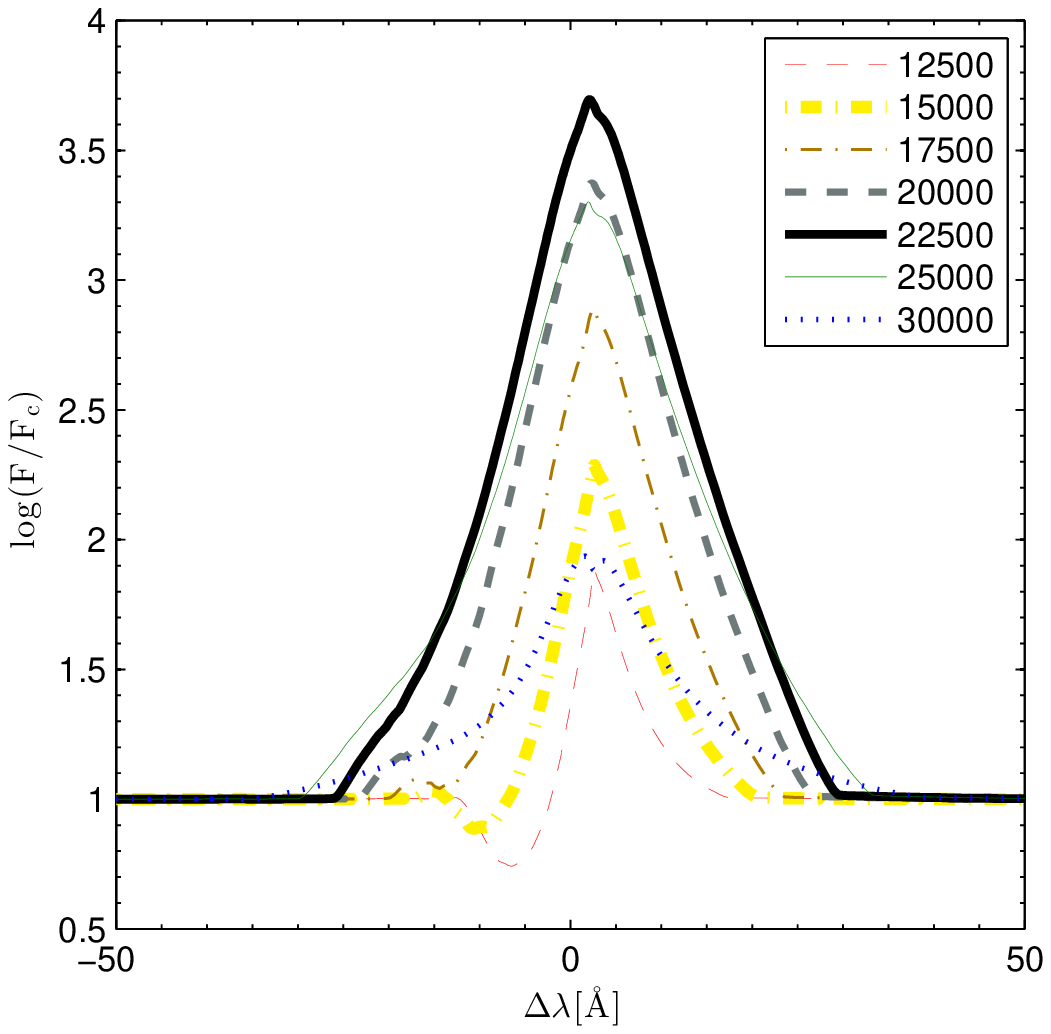}\hspace{\spfactor} \includegraphics{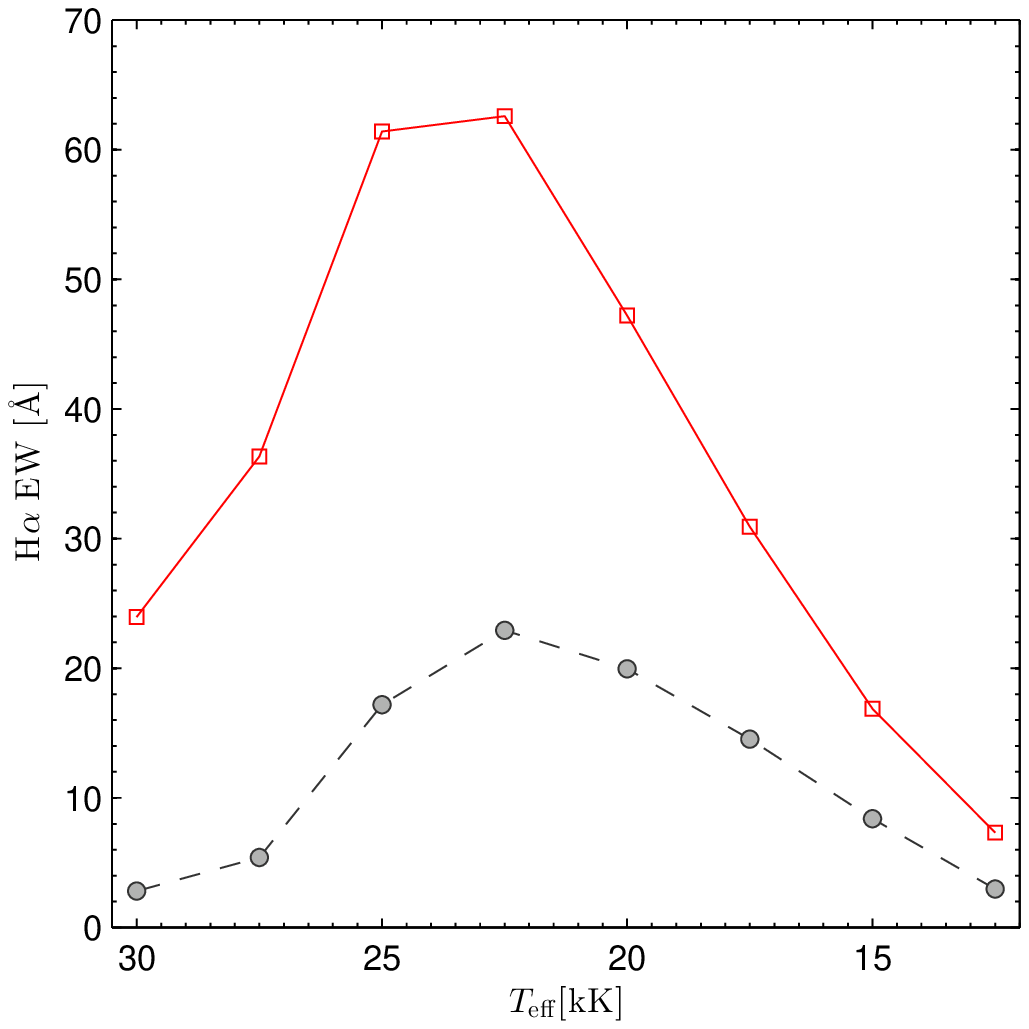}}	
      \caption{\textit{Left:} {\bvp synthetic} H${\alpha}$ line profiles from clumped models with  volume filling factor $f_{\infty}=0.1$. \textit{Right:} H${\alpha}$ line EW as a function of the effective temperature for homogeneous (circles) and clumped (squares) models.}
      \label{fig:HaLP_CL}
    \end{figure*}

      The H$\alpha$ line emission is a $\rho^2$ dependent process and is therefore sensitive to micro-clumping. In order to investigate the potential role of micro-clumping in our model sequences we calculated additional models, identical to the simplistic H+He models but with $f_\infty = 0.1 $, described by the following exponential law:      
      \begin{equation}
	f(r) = f_{\infty} + (1 - f_{\infty}) \rm {exp}(-\varv(r)/\varv_{cl}),
      \end{equation}
     where $\varv_{\rm cl}$ is the velocity at which clumping starts. We have chosen the clumping to start at $\varv_{\rm cl}=20$ km\,s$^{-1}$, just above the sonic point.

     The H$\alpha$ line profiles are presented in the left panel of Fig.~\ref{fig:HaLP_CL}, where it is shown that clumping enables the asymmetry in H$\alpha$ to appear at hotter $T_{\rm eff}$.  When clumping was neglected, the increased level 2 produced asymmetric line profile at 15\,000\,K and 12\,500\,K (cf. Fig.~\ref{fig:HaLP}), whilst in clumped models with $f_{\infty}=0.1$, the asymmetry is already present at 20\,000\,K. The reason for this shift towards higher temperatures is caused by the increased mean density of the micro-clumped winds (in comparison to the mean density of smooth winds), which in turn increases the Ly$\alpha$ optical depth. Consequently, $n_2$ increases at hotter temperatures. The next question is whether micro-clumping may have an effect in terms of the cool versus hot branch sequences? 

    Therefore in the right panel of Fig.~\ref{fig:HaLP_CL} we compare how the H$\alpha$ line EW behaves as a function of $T_{\rm eff}$ for both clumped and unclumped models. It is found that micro-clumping changes the EW dramatically in the hotter models. Also at the bi-stability jump location ($\sim$22\,500\,K) micro-clumping has a dramatic impact on the H$\alpha$ EW. However, micro-clumping progressively plays a lesser role towards the cooler edge of the Bsg regime. The reason for this is that the second level now behaves as a quasi-ground state, i.e., it scales linearly with density $\rho$, and remains rather unaffected, whilst the more drastic effects for hotter models are the result of the $\rho^2$ scaling. Nevertheless, clumping transforms H$\alpha$ from a pure emission line into a P-Cygni line at 15\,000 K. Although micro-clumping is of quantitative relevance in Bsgs (especially around the bi-stability jump), the existence of an  H$\alpha$ EW peak remains present in clumped model sequences. 

    \subsection{The H$\alpha$ optical depth in a micro-clumping approach}

\begin{figure*}
	\centering
	\resizebox{\factor\hsize}{!}
	{\includegraphics{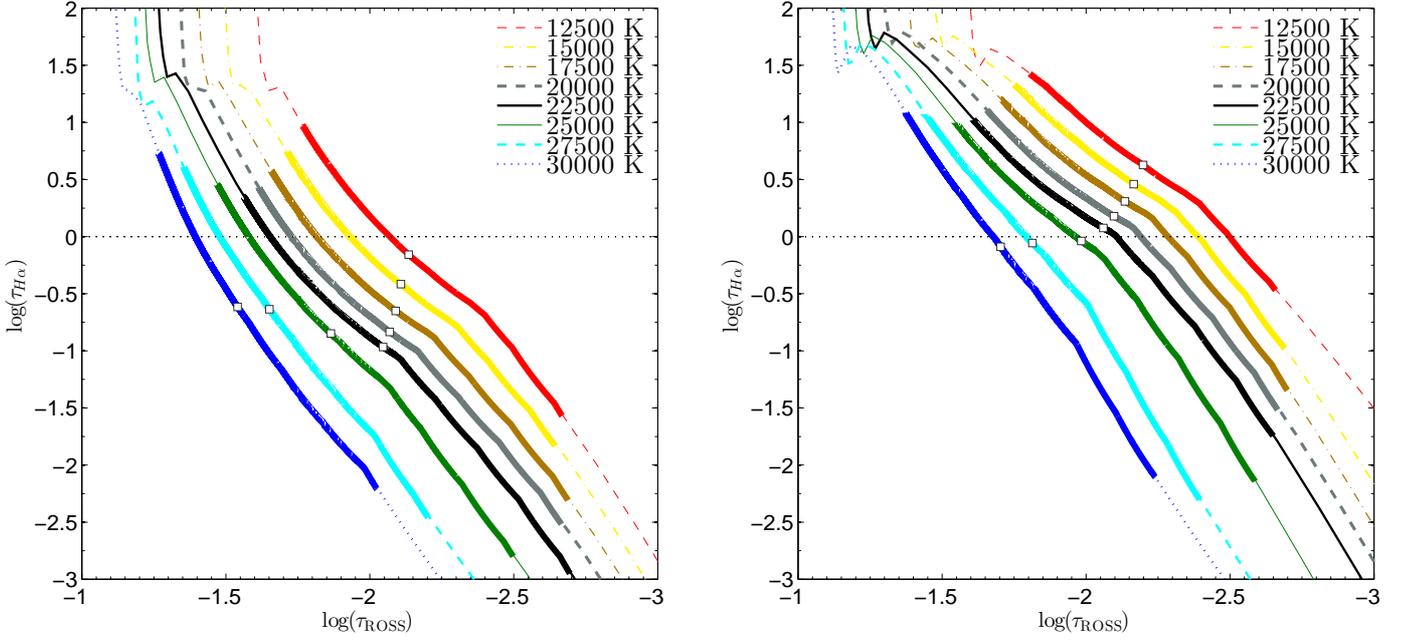}\hspace{\spfactor} \includegraphics{tauHaCL0.1V20kms}}	\caption{ H$\alpha$ Sobolev  optical depth as a function of $\rm \tau_{ross}$ for homogeneous (left) and clumped (right) models. Sites where most of the emergent H$\alpha$ photons originates from are set with {\bvp thick} solid lines. {\bvp White} squares represent the point in which 50\% of the line EW is already formed (see Appendix A).}
	\label{fig:tauHa}
      \end{figure*}

      In Fig.~\ref{fig:tauHa} we compare how the H$\alpha$ optical depth changes with $T_{\rm eff}$ for homogeneous (left) and clumped (right) models. The line-forming region is indicated by thick lines. White squares illustrate at which point 50\% of the line EW is already formed (see Appendix A). According to the left plot most of the H$\alpha$ photons emerge from regions in which the line is optically thin. Interestingly,  when $T_{\rm eff}$ drops from 30\,000 to 22\,500\,K, H$\alpha$ becomes optically thinner. Below 22\,500\,K, the line changes its behaviour and becomes optically thicker with decreasing $T_{\rm eff}$. 
 
      Figure~\ref{fig:tauHa} illustrates that the introduction of      micro-clumping would increase the H$\alpha$ optical depth. It seems that at the bi-stability jump      ($\sim$22\,500\,K) clumping has the largest impact on the      H$\alpha$ optical depth: it increases by an      \textit{order of magnitude} at the location where 50\% of the      line EW is formed (indicated by white squares in      Fig.~\ref{fig:tauHa}). At this point all homogeneous models are      optically thin (left panel), while the clumped models are predominantly optically       thick in H$\alpha$ (right panel).            It appears that the character of H$\alpha$ changes when $T_{\rm eff}$   drops: from an optically thin to an optically thick line (Although, we kept the mass-loss rate {\bvp and \vinf/\vesc} constant in all models). 
      {\bvp Moreover, it is well known from the observations that \vinf/\vesc \,drops from 2.6 to 1.3 across the bi-stability jump \citep{lam95,markova08}. This should produce even sharper differences in  the H$\alpha$ line on both sides of the bi-stability jump, as the higher velocity ratio on the hot side is expected to decrease the mean density and thus H$\alpha$ line optical depth, whilst the lower velocity ratio on the cool side would favour higher optical depth of H$\alpha$ line.}
      It is worth mentioning that the inclusion of clumping in the models     is equivalent to an increase of the mass-loss rate by      factor of $\sqrt{f_{\infty}^{-1}}$ (in comparison to a       homogeneous model)\footnote{We have tested this assumption by	comparing H$\alpha$ optical depth in homogeneous model to a	clumped model with factor of $\sqrt{f_{\infty}^{-1}}$ lower	$\dot{M}$.}. Because of this, Fig.~\ref{fig:tauHa} also      illustrates how the H$\alpha$ line optical depth would change if      the mass-loss rate were increased by factor of      $\sqrt{f_{\infty}^{-1}}$.

      All models discussed here have $\dot{M}=2.33\times10^{-6}\,M_\odot{yr}^{-1}$ which is nearly 3 times larger than the     predicted mass-loss rate for Bsgs around the bi-stability jump.      Thus, our homogeneous models are similar to models with roughly      3 times lower mass-loss rates, but with clumped winds (with      $f_{\infty}=0.1$). Hence, in the context of the predictions, we      expect an increase of the H$\alpha$ line optical depth around      the bi-stability jump analogous to what is shown in      Fig.~\ref{fig:tauHa} (i.e., increased by an order of magnitude      or more if the mass-loss rate is increased by a factor 5; Vink et al. 1999).

    \subsection{Impact of macro-clumping on H$\alpha$}

      Our finding implies that although  changes in the H$\alpha$ optical depth  are largest at the cool branch, the H$\alpha$ line EW is less sensitive to micro-clumping in the coolest models.  The reasons for that are as follows:
      \begin{itemize}
	\item the H$\alpha$ optical depth increases when $T_{\rm eff}$ is reduced.
	\item the Ly$\alpha$ optical depth also increases.  
	\item The second level is prevented to recombine to the ground state.
      \end{itemize}
      As a consequence, the second level behaves as an effective ground state. Thus, on the cool branch the H$\alpha$ line behaves  as an optically thick resonance line  and is therefore  more sensitive to macro-clumping.

     We thus conclude that macro-clumping could play a major role in H$\alpha$ line formation on the cool side of the bi-stability jump.  This could significantly affect empirically derived mass-loss rates. \cite{surlan13} showed that for O-type supergiants macro-clumping may resolve the discrepancy between empirical mass-loss rates derived from H$\alpha$ and P\begin{small}V\end{small} diagnostics \citep[see also][]{sundqvist11}.

  \section{Discussion}
  \label{disc}

    In this theoretical study, we have made progress in our understanding of the H$\alpha$ line over the hot and cool branches around the bi-stability jump. What we have not yet discussed is whether our results have a direct bearing on the reported discrepancies of empirical late Bsg mass-loss rates from model atmosphere analyses and Monte Carlo iron-line driving calculations of \cite{vink00}. In other words, the problem of the general trend that $\dot{M}_{vink} > \dot{M}_{\rm{H}_{\alpha}}$ for B1 and later supergiants.

    Previous investigators, in particular \cite{trundle04,trundle05} and \cite{cro06} argued that wind clumping would make the discrepancy  between $\dot{M}_{\rm H\alpha}$ and $\dot{M}_{\rm Vink}$ worse,  but this conjecture relies on 3 assumptions: (i) that the character of the wind clumps (e.g. optically thin versus optically thick) would remain the same at the bi-stability jump; (ii) that the amount of wind clumping would remain the same; and (iii) that the diagnostic effects of micro-clumping on the H$\alpha$ line would be constant with $T_{\rm eff}$. How likely is it that the wind structure remains the same when the physics changes at the bi-stability jump?

    In the previous sub-section, we noted that our results are in contradiction to those assumptions. The first assumption is also in contradiction to the results of    \cite{prinja10}, who concluded on the basis of Si\begin{small}IV\end{small} lines that the Bsgs winds contain optically {\it thick} clumps, and \cite{Oskinova07} have shown for O star winds that if clumps become optically thick -- and the wind becomes porous -- clumping might in fact {\it under}estimate empirical mass-loss rates.

    Our results show that H$\alpha$ may become optically thick below the bi-stability jump, whilst conversely it may remain optically thin for hotter objects. This might imply that previous empirical mass-loss rates below the bi-stability jump may have been {\it under}estimated, whilst those from hotter stars could be correct or slightly overestimated -- as a result of micro-clumping.    It might also be relevant that the modelled H$\alpha$ lines in the work of \cite{trundle04} and \cite{trundle05} do not reproduce the observed H$\alpha$ line shapes. This suggests that the underlying model used in these analyses might not be correct. Moreover, the ``derived'' values for $\beta$ are much higher than predicted, and may be an artefact of an inappropriate modelling procedure in case macro-clumping would be relevant. It would thus be worthwhile for future investigations if the effects seen in Figs.~\ref{fig:HaLP_CL} and~\ref{fig:tauHa} could indeed explain the reported mass-loss discrepancies.

  \section{Conclusions}
    \label{Conclusion}

    The  behaviour of the H$\alpha$ line over $T_{\rm eff}$ range between 30\,000 and 12\,500\,K might be characterised by the management between two processes.  Whilst the ''rise`` is the result of simple recombination ($n_{3}{^{\uparrow}}$), the ''fall`` is due to the intricate  behaviour of the second level ($n_{2}{^{\uparrow}}$).     As $T_{\rm eff}$ drops below 22\,500\,K, the existence of a cool  branch may be summarised as follows:
    \begin{itemize}
      \item the high Lyman continuum optical depth makes ionisation from the first level unlikely.
      \item Ly$\alpha$ becomes optical thick.
      \item The drain from the second level is suppressed.
      \item At the coolest model level 2 diverges from higher levels, and it  operates like a ground state.
      \item H$\alpha$ changes its character and behaves like a scattering line with a P-Cygni profile.
    \end{itemize}

    \noindent
    During the transition from a recombination to a scattering line, the EW decreases, because recombination lines have a larger (and basically unlimited EW, if the mass-loss rate is increased), whilst a scattering line is confined in its EW as it is dominated by the velocity field. Thus, the EW has to decrease when the line starts to change its character, i.e over the cool branch.

    The qualitatively similar H$\alpha$ behaviour including just H, and H+He only models, and metal-blanketed models suggests that the H$\alpha$ behaviour is not related to He or metal properties. Intriguingly, we also found that the effect of clumping on H$\alpha$ is largest at $T_{\rm eff}$ around  the bi-stability jump and this is not related to the iron ionisation. Furthermore, if the mass-loss rates are increased at the bi-stability jump (as predicted), then the effect (together with clumping) on H$\alpha$ optical depth  would be even stronger. Therefore it is expected that H$\alpha$ changes its character from an optically thin to an optically thick line and the micro-clumping approximation may no longer be justified.

    Although all codes include  the physics explained in this work (except for macro-clumping), it is interesting that independent of model complexity, the H$\alpha$ EW peaks at the location of the bi-stability jump for all our models. This might have consequences for both the physics of the bi-stability mechanism, as well as the derived mass-loss rates from H$\alpha$ line profiles, as objects located below the H$\alpha$ EW peak are predicted to be weaker for a similar mass-loss rate, i.e. higher empirical mass-loss rates are required to reproduce a given H$\alpha$ EW if the star is located at a $T_{\rm eff}$ below the peak. Whether this deeper understanding of H$\alpha$ EW over the bi-stability regime would indeed lead to a resolution of the BSG problem remains to be shown with detailed comparisons of our models and observed H$\alpha$ profiles.
 
    This may also be relevant for magnetic field confinement by magnetic spots that could be induced by the sub-surface convection zone in Bsgs \citep{cantiello09}, as \cite{shultz13} noted that a pair of spots could have remained undetected by current field searches if the higher theoretical mass-loss rates are employed, but that even the largest spots are ruled out if the lower H$\alpha$ rates are correct.

  \begin{acknowledgements}
    We would like to thank Dr. Joachim Puls and an anonymous referee for stimulating discussions and detailed comments on an earlier draft of the manuscript. We acknowledge financial support from the Northern Ireland Department of Culture, Arts, and Leisure (DCAL) and the United Kingdom (UK) Science and Technologies Facilities Council (STFC). We are grateful to Dr. John Hillier for providing the {\sc cmfgen} code  to the astronomical community. 
  \end{acknowledgements}

  \begin{appendix}
    \label{appLFR}
    \vspace{-0.5cm}
    \section{Where in the wind do H$\alpha$ photons originate from?}

      \begin{figure}
	\centering
	\resizebox{\hsize}{!}{\includegraphics{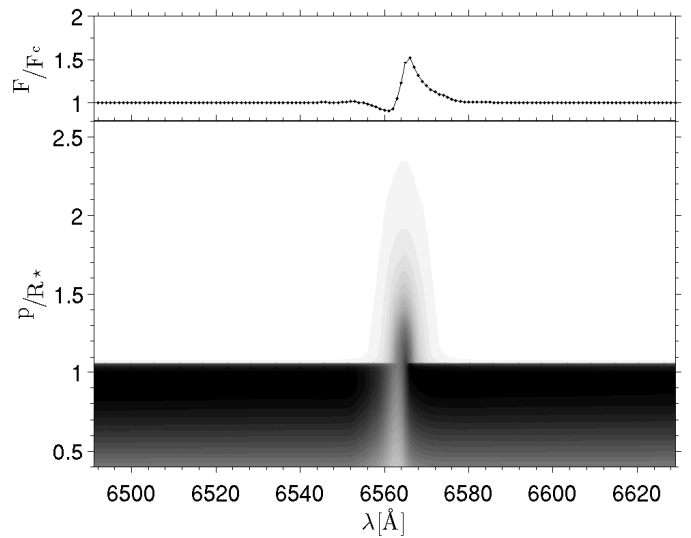}}
	\caption{Model C ($T_{\rm eff}$=12\,500\,K). {\it Bottom}: grey scale plot of the flux like quantity $p\times I(p)$ as a function of impact parameter$p/R_{\star}$, where $R_\star$ is hydrostatic radius. The figure provides the distribution of the emergent intensity around H$\alpha$ from different $p$. \textit{Top:} corresponding  normalised flux in H$\alpha$, directly obtained by integrating $p\times I(p)$ over the range of $p$.}  
	\label{fig:IP12}
      \end{figure} 

      In order to understand the H$\alpha$ line-formation we show in Fig.~\ref{fig:IP12} a typical  distribution of the  emergent  intensity $I(p)$, which is scaled by the impact parameter $p$ \citep[see][for details]{Dess05}. The top panel of the figure  represents the synthetic line profile, which, for each wavelength corresponds to the integral over all $p$ of the scaled intensity $I(p)$ presented at the lower panel. Hence, we are able to identify  the contribution at each $p$ to the total line flux. 

\begin{figure}
	\centering
	\resizebox{\factor\hsize}{!}{\includegraphics{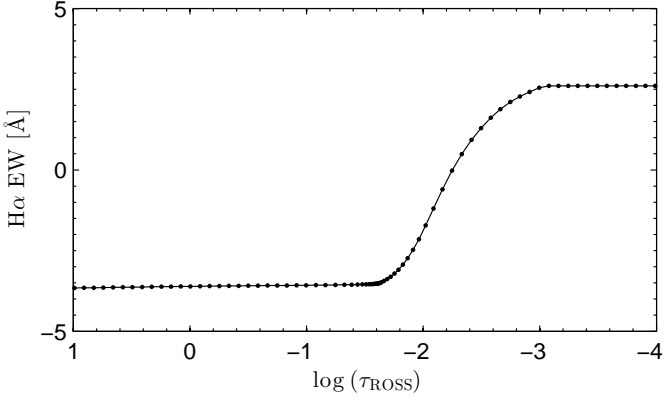}}
	\caption{H$\alpha$ line EW as a function of $\tau_{\rm ross}$. The figure illustrates how the EW changes when outer layers of the star ($p/R_{\rm \star}<~1$) are added.}
	\label{fig:EWvsTAU}
      \end{figure}

      In other words the figure provides  information where in the wind  most of the emergent H$\alpha$ photons originate from. This knowledge enables us  to display the  ''evolution`` of the H$\alpha$ line (EW)  if larger $p$ values are added (or removed) in  Fig.~\ref{fig:EWvsTAU}.  Note that in Fig.~\ref{fig:IP12} the absorption component  emanates for $p/R_{\star}<= \sim1$ (front of the stellar disc), which is in agreement with the conventional mechanism for P~Cygni line formation.

      In  Fig.~\ref{fig:EWvsTAU} we present how the  H$\alpha$ line EW changes when the emergent flux at larger $p/R_{\rm \star}$ (corresponding to $x=r/R_\star$ or $\tau_{\rm ross}$) is added. Note that the H$\alpha$ EW is nearly constant  for log$(\tau_{\rm ross})>\sim-1.5$: in that range the line is in absorption, mainly produced by the wind in front of the stellar disc ($p/R_{\rm \star}<\sim~1)$, i.e the H$\alpha$ photons originate at larger distances. From this figure, we define the H$\alpha$ line-formation region as the region in which  H$\alpha$   changes its  EW from 10\% to 90\%.    Although it is by no means conclusive that the line  forms in this region, most of the H$\alpha$  photons (in the observer's frame) are emitted from this part of the wind, and the behaviour of the line  should  depend on the local conditions in that region. Therefore, we investigated the  H$\alpha$ related  quantities ($(n_3/n_2)$ ratio, $\tau_{\rm Ly\alpha}, \ \tau_{\rm H\alpha})$ at this side of the wind in different models.

  \end{appendix}

\begin{appendix}
  \section{Model atom (sophisticated models)}
    \begin{table}[t] 
    \caption{For each ion the number of full levels, super levels, and bound-bound transitions are provided.}
      \centering 
      \begin{tabular}{l r r r } 
	Ion & Super levels   & Full levels & b-b transitions \\
	\hline
	H  I      &   20 &  30  &   435   \\ He I      &   45 &  69  &   905   \\ 
	He II     &   22 &  30  &   435   \\ C  I      &   81 & 142  &  3426   \\
	C  II     &   40 &  92  &   903   \\
	C  III    &   51 &  84  &   600   \\ C  IV     &   59 &  64  &  1446   \\
	N  I      &   52 & 104  &   855   \\
	N  II     &   45 &  85  &   556   \\ N  III    &   41 &  82  &   578   \\
	N  IV     &   44 &  76  &   497   \\ N  V      &   41 &  49  &   519   \\ 
	O  I 	  &   32 &  161 &  2138   \\
	O  II	  &   54 &  123 &  1375   \\ O  III	  &   88 &  170 &  1762   \\
	O  IV	  &   38 &   78 &   561   \\ O  V	  &   32 &   56 &   314   
\\ O  VI	  &   25 &   31 &   203   \\ Si II	  &    9 &   16 &    37   
\\ Si III	  &   33 &   33 &    92   \\ Si IV	  &   22 &   33 &   185   
\\ P  IV	  &   30 &   90 &   656   \\ P  V      &   16 &   62 &   561   
\\ S  III	  &   24 &   44 &   196   \\ S  IV	  &   51 &  142 &  1504   
\\ S  V  	  &   31 &   98 &   798   \\
	Fe I 	  &    9 &   33 &    47   \\  Fe II	  &  275 &  827 & 23004   
\\ Fe III	  &  104 & 1433 & 57972   \\ Fe IV	  &   74 &  540 & 13071   
\\ Fe V	  &   50 &  220 &  2978   \\ Fe VI	  &   44 &  433 &  8662   \\ Fe 
VII	  &   29 &  153 &  1247   \\ \hline
	\end{tabular}
	\label{table:tabAP} \end{table}
  \end{appendix}
\end{document}